\documentclass[aip,pop,reprint]{revtex4-1}
\usepackage[intlimits]{amsmath}
\usepackage{enumitem}
\usepackage{hyperref}
\usepackage{graphicx}
\usepackage{color}
\usepackage[normalem]{ulem}

\draft 
\allowdisplaybreaks[1]

\newcommand{\vpar}{v_\parallel}
\newcommand{\tderiv}[1]{\frac{\partial #1}{\partial t}}
\newcommand{\pderiv}[2]{\frac{\partial #1}{\partial #2}}
\newcommand{\pgrad}{\nabla^\perp}

\newcommand{\llderiv}{\partial^\parallel}
\newcommand{\pLap}{\Delta^\perp}
\newcommand{\st}[1]{\hbox{\sout{$#1$}}}

\newcommand{\vcspan}{\operatorname{span}}

\begin{document}


\title{JOREK3D: An extension of the JOREK nonlinear MHD code to stellarators} 



\author{N. Nikulsin}
\email{nikita.nikulsin@ipp.mpg.de}
\affiliation{Max Planck Institute for Plasma Physics, Boltzmannstr. 2, 85748 Garching, Germany}
\author{R. Ramasamy}
\affiliation{Max Planck Institute for Plasma Physics, Boltzmannstr. 2, 85748 Garching, Germany}
\affiliation{Max-Planck Princeton Center for Plasma Physics, Princeton, NJ 08544, United States}
\author{M. Hoelzl}
\author{F. Hindenlang}
\author{E. Strumberger}
\author{K. Lackner}
\author{S. G\"unter}
\affiliation{Max Planck Institute for Plasma Physics, Boltzmannstr. 2, 85748 Garching, Germany}
\author{the JOREK Team}
\thanks{See author list of Ref \onlinecite{hoelzl2020the} for full list of team members.}


\date{\today}

\begin{abstract}
Although the basic concept of a stellarator was known since the early days of fusion research, advances in computational technology have enabled the modelling of increasingly complicated devices, leading up to the construction of Wendelstein 7-X, which has recently shown promising results. This recent success has revived interest in the nonlinear 3D MHD modelling of stellarators in order to better understand their performance and operational limits. This paper reports on the extension of the JOREK code to 3D geometries and on the first stellarator simulations carried out with it. The first simple simulations shown here address the classic Wendelstein 7-A stellarator using a reduced MHD model previously derived by us. The results demonstrate that stable full MHD equilibria are preserved in the reduced model: the flux surfaces do not move throughout the simulation, and closely match the flux surfaces of the full MHD equilibrium. Further, both tearing and ballooning modes were simulated, and the linear growth rates measured in JOREK are in reasonable agreement with the growth rates from the CASTOR3D linear MHD code.
\end{abstract}


\maketitle 

\section{Introduction}

The stellarator, having been proposed by Lyman Spitzer in 1951, is one of the oldest plasma confinement concepts potentially applicable as a fusion power plant. However, early stellarators were plagued with problems stemming from neoclassical transport losses, leading to them being largely phased out in favor of tokamaks by the 1970s \cite{boozer2019curl,helander2014theory,freidberg2014ideal}. However, improved mathematical models and increased computational power, which became available by the late 1980s, allowed to overcome the main challenges faced by the stellarator concept. Moreover, the revival of stellarators brought with it a new strategy for fusion research, where numerical modelling drives the development of future machines, as opposed to the traditional strategy, where smaller scale machines had to be built and experimented on before advancing to larger scale machines. The creation of Wendelstein 7-X is one example of the successful application of this new strategy. The advantages are clear: not only is it more cost effective, but it also allows to consider a much wider range of potential machine designs in a much shorter amount of time \cite{boozer2019curl}.
	
Most of the computational developments mentioned above focused on the optimization of stellarator equilibria. While there has been work on nonlinear magnetohydrodynamic (MHD) simulations of stellarators since the 1970s \cite{wakatani1978non}, this area is not as developed as stellarator optimization. Most early studies applied the straight stellarator approximation by neglecting toroidicity \cite{wakatani1978non,wakatani1984pressure,ishii1995nonlinear}, and it wasn't until the 2000s that fully 3D geometries were simulated \cite{strauss2004simulation,mizuguchi2009nonlinear}. At present, several well-known MHD codes exist, with a few of them, including M3D-$C^1$ ~\cite{zhou2021approach}, M3D \cite{strauss2004simulation} and MIPS \cite{sato2017characteristics}, having been extended to stellarators. All three of these codes use full MHD on flux surface aligned grids, except for MIPS, which uses a cylindrical grid. NIMROD, another major tokamak code, is still in the process of being extended to stellarators \cite{sovinec2020development}, although the tokamak capabilities of NIMROD are already enough to simulate a stellarator with an axisymmetric vacuum vessel \cite{schlutt2012numerical}. Also, the FLUXO nonlinear MHD code \cite{hindenlang2021explicit} is applicable to stellarator geometries. However, the stellarator capabilities of these codes have not been used much so far. This paper reports on a similar extension of JOREK, one of the leading nonlinear MHD codes for tokamaks \cite{huysmans2007mhd,czarny2008bezier,hoelzl2020the}, to stellarators. The work consists of two parts: first, a reduced MHD model compatible with three-dimensional geometries was derived by generalizing the ideas of Breslau et al, Izzo et al and Strauss \cite{breslau2009some,izzo1985reduced,strauss1997reduced}, then this model is implemented in the JOREK code and tested on a simple stellarator. The first part of the work has already been published in previous papers \cite{nikulsin2019a,nikulsin2021testing}, and so this paper will present the results of the second part of this effort.

As discussed in our previous papers \cite{nikulsin2019a,nikulsin2021testing}, the magnetic field ansatz and equations in stellarator-capable reduced MHD involve a magnetic scalar potential $\chi$, which represents the part of the magnetic field that is generated by the coils. In the tokamak limit, this potential reduces to $\chi = F_0\phi$, where $\phi$ is the toroidal angle, however a stellarator-capable code using this model will need to allow arbitrary scalar potentials. Fortunately, it is possible to represent an arbitrary $\chi$ analytically: since $\nabla\chi$ is a magnetic field, it must be divergence-free, so $\Delta\chi = 0$. One then needs to find a general solution to the Laplace equation in the toroidal coordinate system $(R,z,\phi)$. This was done by Dommaschk \cite{dommaschk1986representations}, who provides his solution as a sum over harmonics, where any particular solution is determined by the coefficients of these harmonics. Naturally, each harmonic individually satisfies the Laplace equation. In order to determine the coefficients for a particular equilibrium, one needs to first calculate the vacuum field on an $(R,z,\phi)$ grid, which we do using the EXTENDER\_P code \cite{drevlak2005pies}.

Using the Dommaschk potential formulation for $\chi$ in conjunction with non-axisymmetric flux surface aligned grids allows one to simulate stellarators relatively efficiently. The steps to run a stellarator simulation can then be summarized as follows:
\begin{enumerate}[noitemsep,nolistsep]
	\item Calculate an equilibrium for the stellarator in question using the GVEC code \cite{hindenlang2019gvec}
	\item Use the output of GVEC to calculate the contribution to the stellarator's magnetic field from the coils (i.e. the curl-free/vacuum field) with the EXTENDER\_P code
	\item Calculate the coefficients for the Dommaschk representation of the scalar potential from the output of EXTENDER\_P
	\item Build a flux surface aligned grid from the geometry data in the GVEC solution and import it into JOREK
	\item Calculate the initial values for the reduced MHD variables from the GVEC solution
	\item Evolve the system implicitly in time using the stellarator reduced MHD equations in JOREK
\end{enumerate}

The GVEC\cite{hindenlang2019gvec} code mentioned above is a new fixed-boundary 3D MHD equilibrium solver which follows the ideas of the well-established VMEC code \cite{hirshman1983steepest,hirshman1991preconditioned}, assuming nested flux surfaces and using a constraint minimization of the MHD energy. In contrast to VMEC, the radial discretization is based on non-uniform B-Splines of arbitrary order, allowing smooth representation of equilibrium quantities.

The rest of this paper is organized as follows. Section \ref{sec:rmhd} states the reduced MHD equations that will be used throughout the rest of the paper. The same section also discusses the compatibility of full MHD equilibria with reduced MHD and shows that the error introduced by reduced MHD is small. Section \ref{sec:domm} explains how the coefficients of the Dommaschk representation can be calculated from EXTENDER\_P output, while section \ref{sec:init} explains how the initial conditions for the reduced MHD variables can be calculated from the GVEC equilibrium. In section \ref{sec:cnsis}, several simulations of stable equilibria in the Wendelstein 7-A stellarator \cite{w7a1980stabilization} are presented to show that spurious instabilities and problems with maintaining equilibrium do not appear. Finally, section \ref{sec:tearing} shows tearing mode simulations in Wendelstein 7-A and benchmarks the growth rates against the CASTOR3D linear MHD code \cite{strumberger2016castor3d,strumberger2019linear}. Section \ref{sec:ballooning} presents a similar benchmark for ballooning modes in the same device. In addition, appendix \ref{sec:nagrid} briefly discusses the new non-axisymmetric grid feature in JOREK, which allows the stellarator simulations to have flux surface aligned grids. In appendix \ref{sec:linear}, we show that the linearized ideal version of the model presented in section \ref{sec:rmhd} has a self-adjoint operator, and thus ideal perturbations will have real eigenvalues.

\section{Reduced MHD and the residual force}\label{sec:rmhd}

Throughout this paper, we will use the reduced MHD model that was derived in Ref \onlinecite{nikulsin2021testing}. The advantage of reduced MHD is that it allows one to use a larger time step than full MHD by eliminating the shortest timescale in the system; even if implicit time stepping is used, accuracy will deteriorate if the time step is too much larger than the shortest time scale. In the tokamak limit, reduced MHD is well-tested and can accurately model tearing and ballooning modes in a wide range of betas and resistivities \cite{pamela2020extended}. As discussed in Refs \mbox{\onlinecite{nikulsin2019a,nikulsin2021testing}}, the full MHD magnetic field can be written as (no approximations)
\begin{equation}
	\label{eq:fullB}
	\vec B_f = \nabla\chi + \nabla\Psi\times\nabla\chi + \nabla\Omega\times\nabla\psi_v,
\end{equation}
where the first term is the part of the magnetic field generated by the coils, the second is field line bending, and the last term corresponds mostly to field compression, but also adds a small correction to field line bending. Since the vacuum field $\nabla\chi$ must be divergence-free, it can be written in terms of Clebsch potentials: $\nabla\chi = \nabla\psi_v\times\nabla\beta_v$. As discussed in Ref \onlinecite{nikulsin2019a}, one can construct a Clebsch-type coordinate system with coordinates $(\psi_v,\beta_v,\chi)$; this coordinate system will be used further in this section. Expression \eqref{eq:fullB} can be seen as just the plasma-current-induced magnetic field (whose vector potential is $\Psi\nabla\chi + \Omega\nabla\psi_v$, with the $\nabla\beta_V$ component removed by a gauge transform) added to the coil field. The full MHD velocity can be written as (no approximations)
\begin{equation}
	\label{eq:fullv}
	\vec v_f = \frac{\nabla\Phi\times\nabla\chi}{B_v^2} + \vpar\vec B + \pgrad\zeta.
\end{equation}
The terms approximately separate the MHD waves, with the first term containing Alfv\'en waves, the second containing slow magnetosonic waves and the last one containing fast magnetosonic waves. Here, $B_v = |\nabla\chi|$. The reduced model is obtained by setting $\zeta = 0$ and $\Omega = 0$, as well as dropping the component of current perpendicular to $\nabla\chi$ in Ohm's law. In addition to that, we perform a further reduction in this paper by setting $\vpar = 0$. The removal of $\vpar$ decreases the accuracy of the model and narrows the range of scenarios where it is valid \cite{pamela2010influence,finn2019real} (Note that the model tested in Ref \onlinecite{pamela2020extended} has $\vpar \neq 0$.), however the simplified model without $\vpar$ is still applicable to the cases considered here, as will be seen. The resulting model consists of equations (2.9), (2.13), (2.15) and (2.18) from Ref \onlinecite{nikulsin2021testing}, with $\zeta$, $\Omega$ and $\vpar$ zeroed out.
\begin{widetext}
\begin{subequations}
	\label{eq:rmhd}
	\begin{gather}
		\begin{aligned}
			\nabla\cdot\left(\frac{\rho}{B_v^2}\pgrad\tderiv{\Phi}\right) &= \frac{B_v}{2}\left[\frac{\rho}{B_v^2},\frac{(\Phi,\Phi)}{B_v^2}\right] + B_v\left[\frac{\rho\widetilde{\omega}}{B_v^4},\Phi\right] - \nabla\cdot\left(\frac{P}{B_v^2}\pgrad\Phi\right) + \nabla\cdot(\widetilde{j}\vec B) + B_v\left[\frac{1}{B_v^2},p\right] \\
			&+ \nabla\cdot(\mu_\perp\pgrad\widetilde{\omega}) - \pLap(\mu_h\pLap\widetilde{\omega}),
		\end{aligned} \label{eq:phieq} \\
		\tderiv{\rho} = - B_v\left[\frac{\rho}{B_v^2},\Phi\right] + P, \\
		\begin{aligned}
			\tderiv{p} &= - \frac{1}{B_v}\left[p,\Phi\right] - \gamma p B_v\left[\frac{1}{B_v^2},\Phi\right] + \nabla\cdot\left[(\gamma-1)\kappa_\perp\nabla_\perp T + (\gamma-1)\kappa_\parallel\nabla_\parallel T + \frac{p D_\perp}{\rho}\nabla_\perp\rho + \frac{p D_\parallel}{\rho}\nabla_\parallel\rho\right] \\
			&+ (\gamma-1)(S_e + \eta_\mathrm{Ohm}B_v^2\widetilde{j}^2),
		\end{aligned} \\
		\tderiv{\Psi} = \frac{\llderiv\Phi - \left[\Psi,\Phi\right]}{B_v} - \eta(\widetilde{j} - \widetilde{j}_0) + \nabla\cdot(\eta_h\pgrad\widetilde{j}), \\
		\widetilde{j} = \Delta^*\Psi, \label{eq:defj}\\
		\widetilde{\omega} = \pLap\Phi, \label{eq:defom}
	\end{gather}
\end{subequations}
\end{widetext}
where the ideal gas law $p = \rho RT$ applies, and the shorthand $P = \nabla\cdot(D_\perp\nabla_\perp\rho + D_\parallel\nabla_\parallel\rho) + S_\rho$ was used. The operators are defined as follows:
\begin{equation*}
	\begin{gathered}
		\llderiv = B_v^{-1}\nabla\chi\cdot\nabla, \qquad \pgrad = \nabla - B_v^{-1}\nabla\chi\llderiv, \\
		\pLap = \nabla\cdot\pgrad, \qquad \Delta^* = B_v^{-2}\nabla\cdot(B_v^2\pgrad, \\
		\nabla_\parallel = B^{-2}\vec B\vec B\cdot\nabla, \qquad \nabla_\perp = \nabla - \nabla_\parallel
	\end{gathered}
\end{equation*}
In addition, ${[f,g] = B_v^{-1}\nabla\chi\cdot(\nabla f\times\nabla g)}$ is the Poisson bracket of two scalar functions $f$ and $g$, and ${(f,g) = \pgrad f\cdot\pgrad g}$ is the inner product of their gradients. In the reduced model the ansatzes \eqref{eq:fullB} and \eqref{eq:fullv} become
\begin{equation}
	\label{eq:anz}
	\vec B = \nabla\chi + \nabla\Psi\times\nabla\chi, \qquad \vec v = \frac{\nabla\Phi\times\nabla\chi}{B_v^2}.
\end{equation}
In the equations \eqref{eq:rmhd}, $\eta$ is the resistivity, $\mu_\perp$ is a viscosity-like parameter (it is not the same as physical dynamic viscosity, as discussed in Ref \onlinecite{nikulsin2021testing}), $\mu_h$ is the hyperviscosity, $\eta_h$ is the hyperresistivity (artificial dissipation parameters in the $\Phi$ and $\Psi$ equations, respectively, that can help with numerical stabilization), $D_\perp$ is mass diffusion across field lines, $\kappa_\perp$ and $\kappa_\parallel$ are the heat conduction coefficients across and along field lines, respectively, and $S_\rho$ and $S_e$ are the mass and energy sources, respectively. The Ohmic resistivity $\eta_\mathrm{Ohm}$ is a separate parameter, which allows one to neglect part of or all of the resistive contribution to internal energy and simply remove that energy from the system, which can be useful if the resistivity is artificially high.

Finally, there are two auxilliary variables: the normalized current in the $\nabla\chi$ direction $\widetilde{j} = -\nabla\chi\cdot\vec j/B_v^2 = -\nabla\chi\cdot\nabla\times\vec B/(\mu_0 B_v^2)$ and the contravariant $\chi$ component of vorticity $\widetilde{\omega} = -\nabla\chi\cdot\vec\omega = -\nabla\chi\cdot\nabla\times\vec v$, both taken with the opposite sign. The definition equations \eqref{eq:defj} and \eqref{eq:defom} can be obtained by taking the dot product of $\nabla\chi$ with the curl of $\vec B$ and $\vec v$, respectively, and then using the identity $\nabla f\cdot\nabla\times\vec Q = -\nabla\cdot(\nabla f\times\vec Q)$, where $f$ is an arbitrary scalar field and $\vec Q$ is an arbitrary vector field. Instead of simply substituting the definition equations \eqref{eq:defj} and \eqref{eq:defom} into the rest of the equations \eqref{eq:rmhd}, $\widetilde{j}$ and $\widetilde{\omega}$ were treated as separate variables, each with their own degrees of freedom on the finite element grid. These degrees of freedom are evaluated at each time step simultaneously with the degrees of freedom for the other variables by using the definition equations \eqref{eq:defj} and \eqref{eq:defom} alongside with the rest of the equations \eqref{eq:rmhd}. This approach, used in combination with transforming the equations to weak form, allows one to avoid second order derivatives in equations \eqref{eq:rmhd}, except for the hyperviscosity term in equation \eqref{eq:phieq}. Second order derivatives can have discontinuities, as the finite elements in JOREK presently only have $G^1$ continuity \cite{hoelzl2020the}. This also means that if one tries to represent $\chi$ in the finite element basis instead of using the Dommaschk analytical form, one will not be able to avoid discontinuities in the first term on the RHS of equation \eqref{eq:phieq} even after applying integration by parts. In our experience, having discontinuities in the advective terms can decrease numerical accuracy or may lead to numerical instabilities, which was one of the reasons for the existence of $\widetilde{\omega}$ as a separate variable. A recent development in JOREK allows one to use basis functions with higher order $G^n$ continuity, where $n$ is a user-selected parameter \cite{pamela2022a}. This will allow one to eliminate $\widetilde{j}$ and $\widetilde{\omega}$ as separate variables, however it has not been ported to JOREK3D yet.

It is important to note that the $\Phi$ evolution equation \eqref{eq:phieq} was obtained in Ref \onlinecite{nikulsin2021testing} by applying a projection operator to the MHD momentum equation,
\begin{equation}
	\label{eq:mntm}
	\tderiv{}(\rho\vec v) + \nabla\cdot(\rho\vec v\vec v) = \vec j\times\vec B - \nabla p,
\end{equation}
and then inserting the ansatzes \eqref{eq:anz}. The viscosity and hyperviscosity terms are added separately after the projection operator is applied (see Ref \onlinecite{nikulsin2021testing} for more details). The projection operator that produces equation \eqref{eq:phieq} is
\begin{equation}
	\label{eq:phipo}
	\nabla\chi\cdot\nabla\times(B_v^{-2}
\end{equation}
If the $\vpar$ variable is kept in the reduced model, then the following projection operator produces the $\vpar$ evolution equation (not shown here) when applied to equation \eqref{eq:mntm}:
\begin{equation}
	\label{eq:vppo}
	\vec B\cdot
\end{equation}
The $\vpar$ evolution equation is not included in the model \eqref{eq:rmhd} and was not used in any of the simulations presented in this paper, but it will be considered in the equilibrium error discussion, which comprises the remainder of this section.

A natural question that arises when considering reduced MHD is whether or not the reduction preserves equilibria. In other words: if a particular equilibrium solution to the full MHD equations is known, will it also be a solution to the reduced MHD equilibrium equations? As shown in Refs \onlinecite{hoelzl2020the,nikulsin2021testing}, a simple argument involving the Grad-Shafranov equation shows that this is indeed the case in the tokamak limit, where the reduced MHD model \eqref{eq:rmhd} reduces to the model that was already used in the tokamak version of JOREK. However, this does not work for a general stellarator, where a full MHD equilibrium does not exactly satisfy the reduced MHD equilibrium equations, and a residual force arises and contributes to equation \eqref{eq:phieq}. However it can be shown that this contribution is small using an ordering argument.

Let $L_\perp$ be the length scale perpendicular to $\nabla\chi$ and $L_\parallel$ be the length scale along $\nabla\chi$. Then, defining $\lambda \equiv L_\perp/L_\parallel$ as the ordering parameter, the spatial derivatives must satisfy $|\llderiv| \sim \lambda|\pgrad|$. The terms in the full magnetic field \eqref{eq:fullB} are ordered as follows:
\begin{equation*}
	\frac{|\nabla\Psi\times\nabla\chi|}{|\nabla\chi|} \sim |\pgrad\Psi| \sim \lambda,
\end{equation*}
and
\begin{equation*}
	\frac{|\nabla\Omega\times\nabla\psi_v|}{|\nabla\chi|} \sim \frac{F_v}{B_v}|\nabla\Omega| \sim \lambda^2,
\end{equation*}
where $F_v = |\nabla\psi_v|$. Identifying $L_\perp$, $B_v$ and $F_v$ as zeroth-order quantities, $L_\perp = O(1)$, $B_v = O(1)$, $F_v = O(1)$, it follows that $L_\parallel = O(\lambda^{-1})$, $\pgrad = O(1)$, $\llderiv = O(\lambda)$, $\Psi = O(\lambda)$ and $\Omega = O(\lambda^2)$. Meanwhile, the residual force due to the reduction is $\vec f_\mathrm{res} = \nabla p - \vec j\times\vec B = \vec j_f\times\vec B_f - \vec j\times\vec B$, where $\vec B $ is the reduced MHD magnetic field \eqref{eq:anz}, $\vec B_f$ is the full magnetic field \eqref{eq:fullB}, and $\vec j$ and $\vec j_f$ are the curls of the corresponding field divided by $\mu_0$. Note that the residual force is just the difference between the full and reduced MHD Lorentz forces. It arises due to the neglect of the last term of \eqref{eq:fullB} in reduced MHD, and will be present even in the zero $\beta$ limit. Inserting the ansatzes, the following expression is obtained for the residual force:
\begin{equation}
	\label{eq:fres}
	\begin{aligned}
		\vec f_\mathrm{res} &= \frac{1}{\mu_0}[\nabla\times(\nabla\Psi\times\nabla\chi)]\times(\nabla\Omega\times\nabla\psi_v) \\
		&+ \frac{1}{\mu_0}[\nabla\times(\nabla\Omega\times\nabla\psi_v)]\times\nabla\chi \\
		&+ \frac{1}{\mu_0}[\nabla\times(\nabla\Omega\times\nabla\psi_v)]\times(\nabla\Psi\times\nabla\chi) \\
		&+ \frac{1}{\mu_0}[\nabla\times(\nabla\Omega\times\nabla\psi_v)]\times(\nabla\Omega\times\nabla\psi_v)
	\end{aligned}
\end{equation}

After some algebra, the reduced MHD current can be written as
\begin{equation}
	\begin{gathered}
		\vec j = \frac{1}{\mu_0}\nabla\times(\nabla\Psi\times\nabla\chi) = \frac{1}{\mu_0}\Delta^*\Psi\nabla\chi + \vec j^\perp, \\
		\vec j^\perp = \frac{1}{\mu_0}\left(B_v\llderiv\pderiv{\Psi}{q^i}\vec e~^i - g^{kn}\pderiv{\Psi}{q^k}B_v\llderiv g_{ni}\vec e~^i\right),
	\end{gathered}
\end{equation}
where the Einstein summation convention is used, with $k,n \in \{\psi_v,\beta_v,\chi\}$ and $i \in \{\psi_v,\beta_v\}$. Here, $g$ is the metric tensor of the Clebsch-type coordinate system aligned to $\nabla\chi$, which was introduced in Ref \onlinecite{nikulsin2019a}, $q^i$ represents the actual coordinates: $q^i \in \{\psi_v,\beta_v\}$, and $\vec e~^i$ are the contravariant basis vectors: $\vec e~^i \in \{\nabla\psi_v,\nabla\beta_v\}$. With this, the first term in the residual force \eqref{eq:fres} expands to
\begin{equation*}
	-\pderiv{\Omega}{\beta_v}\vec j^\perp\times\nabla\chi - \frac{\Delta^*\Psi}{\mu_0}B_v\llderiv\Omega\nabla\chi\times\vec e_{\beta_v} + B_v\llderiv\Omega\vec j^\perp\times\vec e_{\beta_v}.
\end{equation*}
Since $\vec j^\perp = O(\lambda^2)$, it is easy to see that the first two terms above are $O(\lambda^4)$ and the third term is $O(\lambda^5)$.

The second term in the residual force \eqref{eq:fres} can be expanded as
\begin{equation}
	\label{eq:fres2}
	\begin{aligned}
		&\nabla\psi_v B_v\llderiv(\nabla\Omega\times\nabla\psi_v)_{\psi_v} + \nabla\beta_v B_v\llderiv(\nabla\Omega\times\nabla\psi_v)_{\beta_v} \\
		&- B_v^2\pgrad(\nabla\Omega\times\nabla\psi_v)_\chi
	\end{aligned}
\end{equation}
Note that $\nabla\Omega\times\nabla\psi_v = -(\partial\Omega/\partial\beta_v)\nabla\chi + (\partial\Omega/\partial\chi)\vec e_{\beta_v}/J$, where $\vec e_{\beta_v} = J\nabla\chi\times\nabla\psi_v$ is the covariant basis vector in the $\beta_v$ direction in the Clebsch-type coordinate system aligned to $\nabla\chi$, and $J = [(\nabla\psi_v\times\nabla\beta_v)\cdot\nabla\chi]^{-1} = 1/B_v^2$ is the Jacobian. Furthermore, since $B_v\partial/\partial\chi = \llderiv$, one has $(\nabla\Omega\times\nabla\psi_v)_{\psi_v} = g_{\psi_v\beta_v}B_v\llderiv\Omega$ and $(\nabla\Omega\times\nabla\psi_v)_{\beta_v} = g_{\beta_v\beta_v}B_v\llderiv\Omega$. Thus, the first two terms in \eqref{eq:fres2} are $O(\lambda^4)$ and the third term is $O(\lambda^2)$. However, it is easy to see that the third term in \eqref{eq:fres2} is in the kernel of the projection operator \eqref{eq:phipo}, and so this term will not contribute to the residual force in the $\Phi$ evolution equation \eqref{eq:phieq}. On the other hand, the third term in \eqref{eq:fres2} is not in the kernel of the projection operator \eqref{eq:vppo}, but its image under the operator, which can be written as $B_v^3[\partial\Omega/\partial\beta_v,\Psi]$, will be cancelled by the image of another term, as will be shown below.

The curl of the last term of the full magnetic field \eqref{eq:fullB} can be written as
\begin{equation}
	\label{eq:ltcurr}
	\begin{aligned}
		\nabla\times(\nabla\Omega\times\nabla\psi_v) &= -\nabla\pderiv{\Omega}{\beta_v}\times\nabla\chi + \nabla(B_v\llderiv\Omega)\times\vec e_{\beta_v} \\
		&+ B_v\llderiv\Omega\nabla\times\vec e_{\beta_v}.
	\end{aligned}
\end{equation}
Note that the first term in \eqref{eq:ltcurr} is $O(\lambda^2)$ and the other two terms are $O(\lambda^3)$. Using \eqref{eq:ltcurr}, the third term in the residual force \eqref{eq:fres} becomes
\begin{equation}
	\label{eq:fres3}
	\begin{aligned}
		&-\left(\nabla\pderiv{\Omega}{\beta_v}\times\nabla\chi\right)\times(\nabla\Psi\times\nabla\chi) + O(\lambda^4) \\
		&= \nabla\Psi\cdot\left(\nabla\pderiv{\Omega}{\beta_v}\times\nabla\chi\right)\nabla\chi + O(\lambda^4).
	\end{aligned}
\end{equation}
Terms of order $\lambda^4$ are not written out explicitly in \eqref{eq:fres3}, since there is no need to consider them, as it is already established that there is at least an $O(\lambda^4)$ contribution to equation \eqref{eq:phieq}. As can be seen, the $O(\lambda^3)$ term in \eqref{eq:fres3} will be cancelled by the projection operator \eqref{eq:phipo} and as such will not contribute to the $\Phi$ equation \eqref{eq:phieq}, however it will not be cancelled by the projection operator \eqref{eq:vppo}. Indeed, its image under the operator \eqref{eq:vppo} will be $B_v^3[\Psi,\partial\Omega/\partial\beta_v]$. Note that this image is equal to the negative image of the third term in \eqref{eq:fres2}, which was discussed above. These two images will cancel, and thus the lowest order in which the residual force will contribute to the $\vpar$ evolution equation is $\lambda^4$.

Finally, the last term in the residual force \eqref{eq:fres} is clearly of order $\lambda^4$ or higher, so there is no need to consider it in detail like the other terms. In order to compare the residual force contributions to the other terms in equation \eqref{eq:phieq} and the $\vpar$ evolution equation, some more ordering needs to be done. Consider that the shortest time scale in the reduced system is the Alfv\'en time $\tau_\mathrm{A} \equiv L_\parallel/c_\mathrm{A}$, where the parallel length scale is used because the Alfv\'en wave travels along field lines, and so the time derivative is ordered as $|\partial/\partial t| \sim 1/\tau_\mathrm{A}$. As such, $\partial/\partial t = O(\lambda)$. The $\Phi$ and $\vpar$ terms in the velocity ansatz are then ordered as
\begin{equation*}
	|v_\parallel\vec B|\div\frac{|\nabla\Phi\times\nabla\chi|}{B_v^2} \sim B_v^2\frac{|v_\parallel|}{|\pgrad\Phi|} \sim 1.
\end{equation*}
Assuming that the partial and convective terms in the material derivative are of the same order, $|\partial/\partial t| \sim |\vec v\cdot\nabla|$, one has $\Phi,\vpar = O(\lambda)$. After identifying $\rho = O(1)$ and $p = O(\lambda^2)$, it is clear that the lowest order terms in equations \eqref{eq:mntm} are $O(\lambda^2)$, and both projection operators \eqref{eq:phipo} and \eqref{eq:vppo} are $O(1)$. As such, the lowest order terms in both equation \eqref{eq:phieq} and the $\vpar$ evolution equation are $O(\lambda^2)$. Thus, the residual force contribution to the reduced MHD equations is at least two orders of $\lambda$ higher than the lowest order terms. In section \ref{sec:cnsis}, it will be confirmed with numerical simulations that the residual force is indeed small.

\section{Finding the Dommaschk representation of a scalar potential}\label{sec:domm}

Since $\chi$ is a solution of the Laplace equation in a torus, it can be represented as a summation over toroidal harmonics
\begin{equation}
	\label{eq:chi}
	\chi = F_0\phi + \sum_{n,m} \chi_{n,m},
\end{equation}
where $F_0\phi$ corresponds to a tokamak-like toroidal field, $n$ is the toroidal mode number, $m$ is the poloidal mode number, and each harmonic satisfies the Laplace equation individually: $\Delta\chi_{n,m} = 0$. Dommaschk gives a more explicit representation for $\chi$ \cite{dommaschk1986representations}:
\begin{equation}
	\label{eq:wchi}
	\begin{aligned}
		\widetilde{\chi} = \phi + \sum_{n,m}&\Big[(a_{n,m}\cos n\phi + b_{n,m}\sin n\phi)D_{n,m}(\widetilde{R},\widetilde{z}) \\
		&+ (c_{n,m}\cos n\phi + d_{n,m}\sin n\phi)N_{n,m-1}(\widetilde{R},\widetilde{z})\Big],
	\end{aligned}
\end{equation}
where a tilde denotes normalization: $\chi = F_0\widetilde{\chi}$, $R = R_0\widetilde{R}$ and $z = R_0\widetilde{z}$; the normalization factor $R_0$ is the toroidally averaged radial position of the magnetic axis of the vacuum field. The functions $D_{n,m}$ and $N_{n,m}$ are defined as:
\begin{equation}
	\{D,N\}_{n,m}(\widetilde{R},\widetilde{z}) = \sum_{k=0}^{2k\leq m}\frac{\widetilde{z}^{m-2k}}{(m-2k)!}C^{\{D,N\}}_{n,k}(\widetilde{R}),
\end{equation}
and
\begin{equation}
	\label{eq:cfunc}
	\begin{aligned}
		C^D_{n,k}(\widetilde{R}) &= \sum_{j=0}^k [-(\alpha_j(\alpha^*_{k-n-j}\ln\widetilde{R} + \gamma^*_{k-n-j} - \alpha_{k-n-j}) \\
		&- \gamma_j\alpha^*_{k-n-j} + \alpha_j\beta^*_{k-j})\widetilde{R}^{2j+n} + \beta_j\alpha^*_{k-j}\widetilde{R}^{2j-n}], \\
		C^N_{n,k}(\widetilde{R}) &= \sum_{j=0}^k [(\alpha_j(\alpha_{k-n-j}\ln\widetilde{R} + \gamma_{k-n-j}) \\
		&- \gamma_j\alpha_{k-n-j} + \alpha_j\beta_{k-j})\widetilde{R}^{2j+n} - \beta_j\alpha_{k-j}\widetilde{R}^{2j-n}].
	\end{aligned}
\end{equation}
The coefficients $\alpha_i$, $\beta_i$ and $\gamma_i$ are defined as
\begin{equation}
	\begin{gathered}
		\alpha_i = \frac{(-1)^i}{2^{2i+n}\Gamma(n+i+1)\Gamma(i+1)}, \quad \alpha^*_i = (2i+n)\alpha_i, \\
		\beta_i = \frac{\Gamma(n-i)}{2^{2i-n+1}\Gamma(i+1)}, \quad \beta^*_i = (2i-n)\beta_i, \\
		\gamma_i = \frac{\alpha_i}{2}\sum_{j=1}^i\left(\frac{1}{j} + \frac{1}{n+j}\right), \quad \gamma^*_i = (2i+n)\gamma_i.
	\end{gathered}
\end{equation}
Although not written out explicitly, it can be seen that the coefficients also depend on $n$, the toroidal mode number of the $D$ or $N$ function that is being evaluated. The expressions above are only well defined if the following conditions on $i$ and $n$ are met: $i \geq 0$ for $\alpha_i$ and $\alpha^*_i$, $i \geq 0$ and $n > i$ for $\beta_i$ and $\beta^*_i$, and $i > 0$ for $\gamma_i$ and $\gamma^*_i$. Otherwise, the corresponding coefficient and its starred version are zero. Finally, the coefficients $a_{n,m}$, $b_{n,m}$, $c_{n,m}$ and $d_{n,m}$ in equation \eqref{eq:wchi} are what determines a particular configuration and must be calculated from the EXTENDER\_P output.

Note that, since the harmonics $\chi_{n,m}$ are given analytically, the property that $\Delta\chi_{n,m} = 0$ is satisfied exactly. This is an important advantage of using the Dommaschk representation for $\chi$ instead of the finite element representation (see appendix \ref{sec:nagrid}), as it guarantees that the divergence-free condition on the magnetic field will be satisfied to machine precision. The second advantage is that $\chi$ and its derivatives are smooth.

EXTENDER\_P provides the values of the three cylindrical components of the vacuum magnetic field, which will be referred to as $\vec B_\mathrm{E}$, on an $(R,z,\phi)$ grid. Setting $\nabla\chi = \vec B_\mathrm{E}$ and considering the $\phi$ component, $B_{\mathrm{E},\phi} = \widehat{\phi}\cdot\vec B_\mathrm{E} = R^{-1}\partial\chi/\partial\phi$, one has:
\begin{equation}
	\label{eq:bphi}
	\begin{alignedat}{3}
		&\frac{R_0}{F_0}B_{\mathrm{E},\phi} =&& \widetilde{R}^{-1}\frac{\partial\widetilde{\chi}}{\partial\phi} = \widetilde{R}^{-1} \\
		&+ \widetilde{R}^{-1}\sum_{n,m}&&n\Big[(-a_{n,m}\sin n\phi + b_{n,m}\cos n\phi)D_{n,m}(\widetilde{R},\widetilde{z}) \\
		& &&+ (-c_{n,m}\sin n\phi + d_{n,m}\cos n\phi)N_{n,m-1}(\widetilde{R},\widetilde{z})\Big].
	\end{alignedat}
\end{equation}
We will make use of the properties (from Ref \onlinecite{dommaschk1986representations}, equations (10) and (11)) $D_{n,m}|_{\widetilde{R}=1} = \widetilde{z}^m/m!$ and $N_{n,m}|_{\widetilde{R}=1} = 0$. Evaluating equation \eqref{eq:bphi} at $\widetilde{R}=1$ gives:
\begin{equation}
	\left.\frac{R_0}{F_0}B_{\mathrm{E},\phi}\right|_{\widetilde{R}=1} = 1 + \sum_{n,m} n(-a_{n,m}\sin n\phi + b_{n,m}\cos n\phi)\frac{\widetilde{z}^m}{m!}.
\end{equation}
If one also evaluates at $\widetilde{z}=0$ and integrates over $\phi$, $F_0$ can be calculated:
\begin{equation}
	F_0 = \frac{R_0}{2\pi}\int_0^{2\pi} \left.B_{\mathrm{E},\phi}\right|_{\widetilde{R}=1,\widetilde{z}=0} d\phi.
\end{equation}
To calculate the coefficients $a_{n,m}$ and $b_{n,m}$, one must first multiply by either $\sin n\phi$ or $\cos n\phi$ and then use the orthogonality property of trigonometric functions:
\begin{equation}
	\label{eq:lsumz}
	\begin{gathered}
		-n\sum_m a_{n,m}\frac{\widetilde{z}^m}{m!} = \frac{R_0}{F_0\pi}\int_0^{2\pi}\left.B_{\mathrm{E},\phi}\right|_{\widetilde{R}=1}\sin n\phi~d\phi, \\
		n\sum_m b_{n,m}\frac{\widetilde{z}^m}{m!} = \frac{R_0}{F_0\pi}\int_0^{2\pi}\left.B_{\mathrm{E},\phi}\right|_{\widetilde{R}=1}\cos n\phi~d\phi.
	\end{gathered}
\end{equation}

\begin{figure}
	\centering
	\includegraphics[scale=0.6]{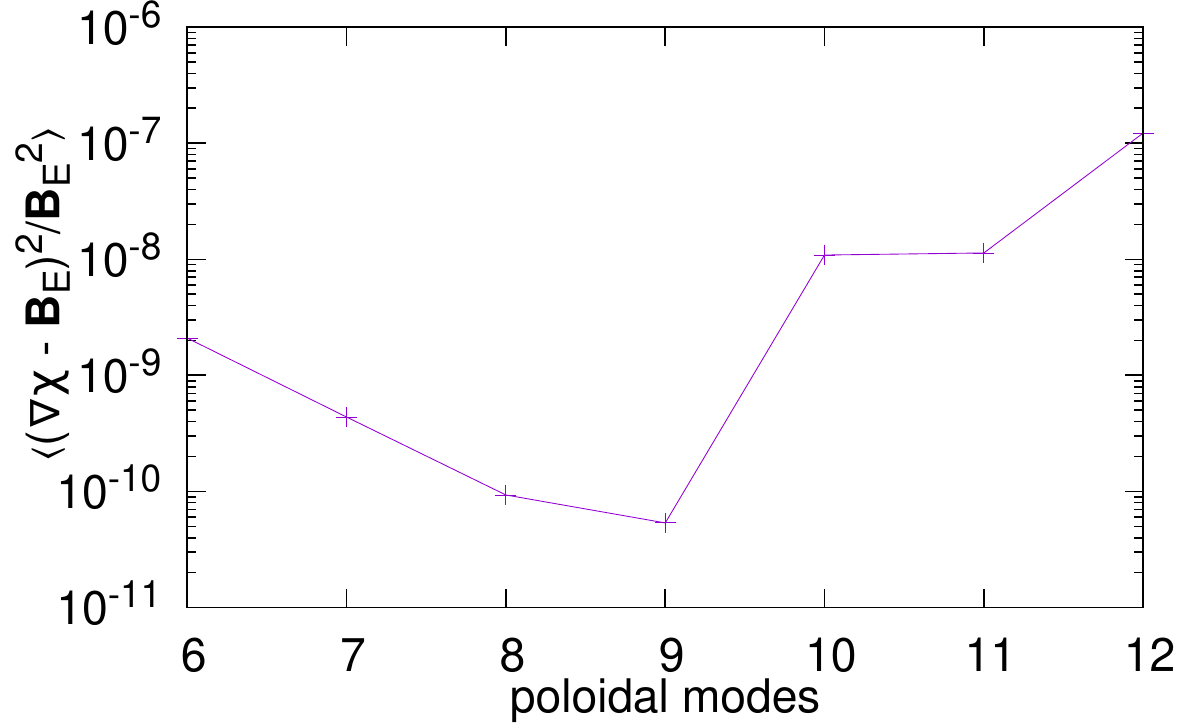}
	\caption{The volume-averaged squared relative error of the Dommaschk potential representation $\langle(\nabla\chi - \vec B_\mathrm{E})^2/B_\mathrm{E}^2\rangle$ as a function of the number of poloidal modes $M$. The values shown in this plot were calculated using a Python implementation of Dommaschk potentials based on the one written by Paul Huslage for the BOUT++ code \cite{dudson2009bout,dudson2016verification}.}
	\label{fig:dpacc}
\end{figure}

The number of terms $M$ in the summations over $m$ in equations \eqref{eq:lsumz} that is necessary to accurately represent the magnetic field is usually less than the number of poloidal modes used in the GVEC equilibrium. In practice, it is best to scan through different values of $M$, starting with the number of poloidal modes and decreasing from there, while trying to minimize the error in $\nabla\chi$ as compared to $\vec B_\mathrm{E}$. Note that using higher values of $M$ than necessary can lead to higher errors away from the $\widetilde{R}=1$ surface due to overfitting, as the integration in equations \eqref{eq:acoef}, \eqref{eq:bcoef}, \eqref{eq:ccoef} and \eqref{eq:dcoef}, which will be derived shortly, is only over the $\widetilde{R}=1$ surface. Figure \ref{fig:dpacc} shows the volume-averaged relative squared error of the Dommaschk potential representation as a function of the number $M$ of poloidal modes kept in a Wendelstein 7-A equilibrium with $\beta = 2.3\cdot 10^{-3}~\%$ (see section \ref{sec:cnsis} for more details about this equilibrium).

\begin{figure}[b]
	\centering
	\includegraphics[scale=0.75]{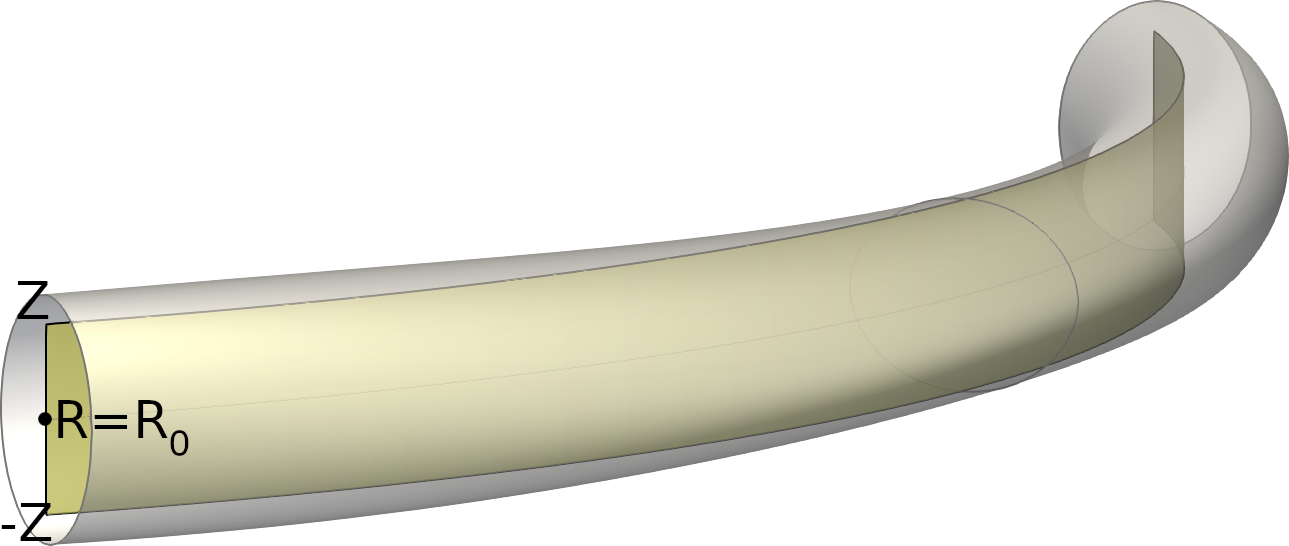}
	\caption{The surface of integration for calculating Dommaschk potential coefficients, $R = R_0$, $-Z \leq z \leq Z$, in one field period of Wendelstein 7-A.}
	\label{fig:isurf}
\end{figure}

One can convert equations \eqref{eq:lsumz} into two linear algebraic systems with triangular matrices by changing the variable to $z' = \widetilde{z}/Z$ and, after multiplying both equations by a Legendre polynomial $P_i(z')$, integrating from -1 to 1. Here, $Z$ is determined as follows. In each poloidal plane at $\widetilde{R}=1$, $\widetilde{z}\in [-\widetilde{z}_-(\phi),\widetilde{z}_+(\phi)]$, so $Z < \min_{\phi} \{\widetilde{z}_-(\phi),\widetilde{z}_+(\phi)\}$. The value of $Z$ is chosen to be slightly smaller than the minimum to avoid using the components of $\vec B_\mathrm{E}$ close to the boundary, where the output of EXTENDER\_P can be less accurate. There is some freedom in choosing the specific value of $Z$, and it may take some trial and error to find the best value. As an example, Figure \ref{fig:isurf} shows a segment of the surface of integration in one field period of the Wendelstein 7-A equilibria described in section \ref{sec:cnsis}.

The Legendre polynomials are orthogonal to monomials of lower order than the polynomial, since a monomial $z^m$ can be expanded exactly in the Legendre polynomial basis of the same order $m$:
\begin{equation*}
	z^m = \sum_{i=0}^m C_{m,i}P_i(z).
\end{equation*}

Using the orthogonality property discussed above, one has, starting with $i=M$ and descending to $i=0$, the following linear algebraic system for $a_{n,m}$:
\begin{equation}
	\label{eq:acoef}
	\begin{aligned}
		&\sum_{m=i}^M\frac{a_{n,m}}{m!}Z^m\langle z^m,P_i(z)\rangle \\
		&= \frac{-R_0}{nF_0\pi}\int_0^{2\pi}\int_{-1}^1\left.B_{\mathrm{E},\phi}\right|_{\widetilde{R}=1,\widetilde{z}=Zz'}P_i(z')\sin n\phi~dz'~d\phi, \\
		&i = M, M-1, ..., 0,
		\end{aligned}
\end{equation}
where $\langle z^i,P_j(z)\rangle = \int_{-1}^1 z^i P_j(z)dz$. Similarly, for the coefficients $b_{n,m}$, one has the following linear algebraic system:
\begin{equation}
	\label{eq:bcoef}
	\begin{aligned}
		&\sum_{m=i}^M\frac{b_{n,m}}{m!}Z^m\langle z^m,P_i(z)\rangle \\
		&= \frac{R_0}{nF_0\pi}\int_0^{2\pi}\int_{-1}^1\left.B_{\mathrm{E},\phi}\right|_{\widetilde{R}=1,\widetilde{z}=Zz'}P_i(z')\cos n\phi~dz'~d\phi, \\
		&i = M, M-1, ..., 0.
	\end{aligned}
\end{equation}
As can be seen, both of these systems of equations have triangular matrices.

At this point, the equations for the coefficients $c_{n,m}$ and $d_{n,m}$ have yet to be determined. Consider now the $R$ component of $\vec B_\mathrm{E}$: $B_{\mathrm{E},R} = \widehat{R}\cdot\vec B_\mathrm{E} = \partial\chi/\partial R$. One has:
\begin{equation}
	\begin{aligned}
		\frac{R_0}{F_0}B_{\mathrm{E},R} = \frac{\partial\widetilde{\chi}}{\partial\widetilde{R}} &= \sum_{n,m}\Bigg[(a_{n,m}\cos n\phi + b_{n,m}\sin n\phi)\frac{\partial D_{n,m}}{\partial\widetilde{R}} \\
		&+ (c_{n,m}\cos n\phi + d_{n,m}\sin n\phi)\frac{\partial N_{n,m-1}}{\partial\widetilde{R}}\Bigg].
	\end{aligned}
\end{equation}
Again, evaluating at $\widetilde{R}=1$ and using the properties $\partial D_{n,m}/\partial\widetilde{R}|_{\widetilde{R}=1} = 0$ and $\partial N_{n,m}/\partial\widetilde{R}|_{\widetilde{R}=1} = \widetilde{z}^m/m!$ (also from Ref \onlinecite{dommaschk1986representations}, equations (10) and (11)), one has:
\begin{equation}
	\left.\frac{R_0}{F_0}B_{\mathrm{E},R}\right|_{\widetilde{R}=1} = \sum_{n,m} (c_{n,m}\cos n\phi + d_{n,m}\sin n\phi)\frac{\widetilde{z}^{m-1}}{(m-1)!}.
\end{equation}
From here, it is straightforward to follow the same steps as for $a_{n,m}$ and $b_{n,m}$, obtaining the following linear algebraic systems for $c_{n,m}$:
\begin{equation}
	\label{eq:ccoef}
	\begin{aligned}
		&\sum_{m=i}^{M-1}\frac{c_{n,m+1}}{m!}Z^m\langle z^m,P_i(z)\rangle \\
		&= \frac{R_0}{F_0\pi}\int_0^{2\pi}\int_{-1}^1\left.B_{\mathrm{E},R}\right|_{\widetilde{R}=1,\widetilde{z}=Zz'}P_i(z')\cos n\phi~dz'~d\phi, \\
		&i = M-1, M-2, ..., 0,
	\end{aligned}
\end{equation}
and for $d_{n,m}$:
\begin{equation}
	\label{eq:dcoef}
	\begin{aligned}
		&\sum_{m=i}^{M-1}\frac{d_{n,m+1}}{m!}Z^m\langle z^m,P_i(z)\rangle \\
		&= \frac{R_0}{F_0\pi}\int_0^{2\pi}\int_{-1}^1\left.B_{\mathrm{E},R}\right|_{\widetilde{R}=1,\widetilde{z}=Zz'}P_i(z')\sin n\phi~dz'~d\phi, \\
		&i = M-1, M-2, ..., 0.
	\end{aligned}
\end{equation}
Note that there are only $M$ equations in each system for the unknowns $c_{n,1},\ldots,c_{n,M}$ and $d_{n,1},\ldots,d_{n,M}$ because $N_{n,-1}$ is not defined, and so terms with $c_{n,0}$ and $d_{n,0}$ are not included in the sum \eqref{eq:wchi}.

The only coefficients for which a system of equations has not yet been obtained are $a_{0,m}$ (there are no $b_{0,m}$ coefficients since $\sin 0 = 0$). These coefficients cannot be obtained from the system \eqref{eq:acoef} since the matrices of this system are singular when $n=0$. To get a solvable system, one must use the $z$ component of $\vec B_\mathrm{E}$:
\begin{equation}
	\label{eq:a0coef}
	\begin{aligned}
		\frac{R_0}{F_0}B_{\mathrm{E},z} = \frac{\partial\widetilde{\chi}}{\partial\widetilde{z}} &= \sum_{n,m}\Bigg[(a_{n,m}\cos n\phi + b_{n,m}\sin n\phi)\frac{\partial D_{n,m}}{\partial\widetilde{z}} \\
		&+ (c_{n,m}\cos n\phi + d_{n,m}\sin n\phi)\frac{\partial N_{n,m-1}}{\partial\widetilde{z}}\Bigg].
	\end{aligned}
\end{equation}
Evaluating at $\widetilde{R} = 1$ using the properties (from Ref \onlinecite{dommaschk1986representations}, equations (10) and (11)) $D_{n,m}|_{\widetilde{R}=1} = \widetilde{z}^m/m!$ and $N_{n,m}|_{\widetilde{R}=1} = 0$ after differentiating by $\widetilde{z}$ gives:
\begin{equation}
	\left.\frac{R_0}{F_0}B_{\mathrm{E},z}\right|_{\widetilde{R} = 1} = \sum_{n,m} (a_{n,m}\cos n\phi + b_{n,m}\sin n\phi)\frac{\widetilde{z}^{m-1}}{(m-1)!}
\end{equation}
Integrating over $\phi$ leaves only the $n=0$ term in the sum, as all others are harmonic:
\begin{equation}
	\sum_m a_{0,m}\frac{\widetilde{z}^{m-1}}{(m-1)!} = \frac{R_0}{2F_0\pi}\int_0^{2\pi}\left.B_{\mathrm{E},z}\right|_{\widetilde{R}=1}d\phi.
\end{equation}
To finalize the derivation, we multiply the equation by a Legendre polynomial $P_i(z')$ and integrate from -1 to 1. Starting from $i=M-1$ and descending to $i=0$, the system of equations is
\begin{equation}
	\begin{aligned}
		&\sum_{m=i}^{M-1} \frac{a_{0,m+1}}{m!}Z^m\langle z^m, P_i(z)\rangle \\
		&= \frac{R_0}{2F_0\pi}\int_0^{2\pi}\int_{-1}^1\left.B_{\mathrm{E},z}\right|_{\widetilde{R}=1,z=Zz'}P_i(z')dz'd\phi, \\
		&i = M-1, M-2, ..., 0.
	\end{aligned}
\end{equation}
Just as in the case of systems \eqref{eq:ccoef} and \eqref{eq:dcoef}, there are only $M$ equations for the unknowns $a_{0,1},\ldots,a_{0,M}$. This is because $D_{0,0} = 1$, and so $a_{0,0}$ is an additive constant in the scalar potential, which has no effect on the vacuum magnetic field  \cite{dommaschk1986representations}.

The linear algebraic systems of equations \eqref{eq:acoef}, \eqref{eq:bcoef}, \eqref{eq:ccoef}, \eqref{eq:dcoef} and \eqref{eq:a0coef} are solved in a Python script using the NumPy library \cite{harris2020array}. The solution is then written out to a Fortran namelist file, which can be read by JOREK. When evaluating the Dommaschk potential and its derivatives at any particular point, JOREK will then use the analytical representation \eqref{eq:wchi}.

\section{Determining initial conditions from the GVEC solution}\label{sec:init}

As was mentioned in the section \ref{sec:rmhd}, although $\widetilde{j}$ and $\Psi$ are related by $\widetilde{j} = \Delta^*\Psi$, $\widetilde{j}$ is stored as a separate variable in finite element representation for numerical purposes. It makes sense to first calculate the initial condition for $\widetilde{j}$ from the GVEC data, and then calculate $\Psi_0$ from $\widetilde{j}_0$ using equation \eqref{eq:defj} at $t=0$:
\begin{equation}
	\label{eq:Psi0eq}
	\Delta^*\Psi_0 = \widetilde{j}_0.
\end{equation}
Here, the subscript 0 refers to the fact that $\Psi_0$ and $\widetilde{j}_0$ are the initial values of $\Psi$ and $\widetilde{j}$.

The equilibrium magnetic field provided by the GVEC solution will be referred to as $\vec B_\mathrm{GVEC}$. Since GVEC works with full MHD, one needs to consider the full MHD ansatz, as given by \onlinecite{nikulsin2021testing}, when working with $\vec B_\mathrm{GVEC}$:
\begin{equation*}
	\vec B_\mathrm{GVEC} = \nabla\chi + \nabla\Psi_0\times\nabla\chi + \nabla\Omega_0\times\nabla\psi_v.
\end{equation*}
Taking the curl of the above equation and dotting it with $\nabla\chi$, one has, after some algebra:
\begin{equation*}
	\begin{aligned}
		j^\chi_\mathrm{GVEC} = \nabla\chi\cdot\nabla\times\vec B_\mathrm{GVEC} = &-\nabla\cdot(B_v^2\pgrad\Psi_0) \\
		&+ \nabla\cdot(B_v\llderiv\Omega_0\nabla\psi_v).
	\end{aligned}
\end{equation*}
Using the same ordering as in section \ref{sec:rmhd}, where $B_v = O(1)$, $\Psi = O(\lambda)$, $\Omega = O(\lambda^2)$ and $\llderiv = O(\lambda)$, it can be seen that the first term is $O(\lambda)$ and second term is $O(\lambda^3)$. Thus, the second term can be neglected, due to being two orders of $\lambda$ higher than the first term. This significantly simplifies the calculation, as now one can just set $\widetilde{j}_0 = -j^\chi_\mathrm{GVEC}/B_v^2 = -\nabla\chi\cdot\nabla\times\vec B_\mathrm{GVEC}/B_v^2$.

Having determined $\widetilde{j}_0$, it remains to solve the differential equation \eqref{eq:Psi0eq} for $\Psi_0$. First, however, one needs to determine the boundary condition on $\Psi$. Note that $\Psi$ is not constant on flux surfaces in stellarator geometry, unlike the tokamak situation. When running a fixed boundary simulation, as done in this paper, it is usually assumed that the plasma is surrounded by a perfect conductor, so the magnetic field at the boundary does not have a normal component: $\vec n\cdot\vec B = 0$. In the reduced MHD model, this means that $\Psi$ has to satisfy $\vec n\cdot(\nabla\Psi\times\nabla\chi) = -\vec n\cdot\nabla\chi$ at all times. This is a nonhomogeneous linear differential equation which must be solved on the boundary of the torus; the solution to this differential equation then provides a nonhomogeneous Dirichlet boundary condition for equation \eqref{eq:Psi0eq}. Note that the kernel of the differential operator in the boundary equation is quite large, consisting of all functions $f(\chi)$. Using a flux surface aligned coordinate system $(\psi,\theta,\phi)$, where $\psi$ is a flux surface label, $\theta$ is the GVEC poloidal angle, which, like in VMEC, is constructed for each particular equilibrium in the course of minimization, and $\phi$ is the geometric toroidal angle, the boundary equation becomes:
\begin{equation}
	\frac{\partial\Psi}{\partial\theta}\frac{\partial\chi}{\partial\phi} - \frac{\partial\Psi}{\partial\phi}\frac{\partial\chi}{\partial\theta} = -J\nabla\psi\cdot\nabla\chi,
\end{equation}
where $J = [\nabla\psi\cdot(\nabla\theta\times\nabla\phi)]^{-1}$ is the Jacobian. Note that $(\psi,\theta,\phi)$ is not a straight field line coordinate system. However, solving the equation in this form is numerically difficult because one cannot easily separate the kernel and remove it from the solution space. To do so, one must switch to a coordinate system where $\chi$ is one of the coordinates. It is best to switch out $\phi$ for $\chi$, since a stellarator must have a nonvanishing toroidal component to its vacuum field (c.f. the $F_0\phi$ term in equation \eqref{eq:chi}), so $\partial\chi/\partial\phi$ is nonvanishing and the Jacobian of the new coordinates is nowhere singular. The boundary equation in $(\psi,\theta,\chi)$ coordinates is
\begin{equation}
	\label{eq:bndeq}
	\frac{\partial\Psi}{\partial\theta} = -J'\nabla\psi\cdot\nabla\chi,
\end{equation}
where $J' = [\nabla\psi\cdot(\nabla\theta\times\nabla\chi)]^{-1}$ is the new Jacobian. It is easy to solve this equation in JOREK. Due to the JOREK grid for our applications here being flux surface aligned, the element local coordinates $s$ and $t$ (see appendix \ref{sec:nagrid}) can be related to the coordinates $\psi$ and $\theta$ as $\psi = s$, $\theta = 2\pi(t+i_\mathrm{bnd\_elm})/N_\mathrm{bnd\_elm}$, where $i_\mathrm{bnd\_elm}$ is the zero-based index of the current boundary element and $N_\mathrm{bnd\_elm}$ is the total number of boundary elements. Finally, the $\chi$ coordinate is given by the Dommaschk representation \eqref{eq:wchi}. The solution space in which the solution to equation \eqref{eq:bndeq} is searched for can now be represented as:
\begin{equation}
	\begin{aligned}
		\mathcal{V}_\mathrm{sol} = \vcspan [&\{\cos m\theta, \sin m\theta | m=1,...,m_\mathrm{pol}\}\\
		\times&\{1, \cos nN_\mathrm{p}\widetilde{\chi}, \sin nN_\mathrm{p}\widetilde{\chi} | n=1,...,n_\mathrm{tor}\}],
	\end{aligned}
\end{equation}
where $N_\mathrm{cp}$ is defined in Appendix \ref{sec:nagrid} and $\widetilde{\chi} = \chi/F_0$. Excluding the $m=0$ mode removes the kernel of the differential operator of equation \eqref{eq:bndeq} from $\mathcal{V}_\mathrm{sol}$, and the equation can then be solved using the standard Fourier-Galerkin method. The solution obtained this way is then projected back onto the JOREK finite element basis and written to the boundary nodes. Finally, equation \eqref{eq:Psi0eq} is solved by splitting $\Psi_0 = \Psi_{0,\mathrm{i}} + \Psi_\mathrm{b}$, where $\Psi_\mathrm{b}$ is the solution to equation \eqref{eq:bndeq} and thus satisfies the nonhomogeneous Dirichlet boundary condition, while $\Psi_{0,\mathrm{i}}$ is an unknown function which is zero at the boundary. The solution $\Psi_{0,\mathrm{i}}$ is then found using the standard JOREK solver with homogeneous Dirichlet boundary conditions. When $\Psi$ is evolved in time, JOREK will solve for the increment $\delta\Psi$ at each time step, which must also be zero at the boundary (and thus can also be obtained using the standard solver with homogeneous Dirichlet boundary conditions), so that the nonhomogeneous boundary condition continues to be satisfied for the total $\Psi$.

The last step is determining an initial condition for temperature, which is almost trivial. The GVEC solution provides a pressure profile $p_\mathrm{GVEC}$, which must simply be converted to JOREK units and divided by the initial density profile $\rho_0$. In all of the stellarator simulations presented in this paper, the initial density is taken to be constant for simplicity, which corresponds to $\rho_0 = 1$ in JOREK units.

\section{A consistency check for the stellarator model}\label{sec:cnsis}

After having derived and implemented the stellarator model, it remains to validate it for stellarators, showing that it does work. However, before proceeding to more complicated cases, a set of initial tests must be done using stable equilibria to demonstrate that the model is indeed consistent, the error due to neglecting of fourth-order terms in the Lorentz force, which was discussed in section \ref{sec:rmhd}, is small, and no significant change is observed in the stable cases after simulating them for some time.

\begin{figure}
	\centering
	\includegraphics[scale=0.75]{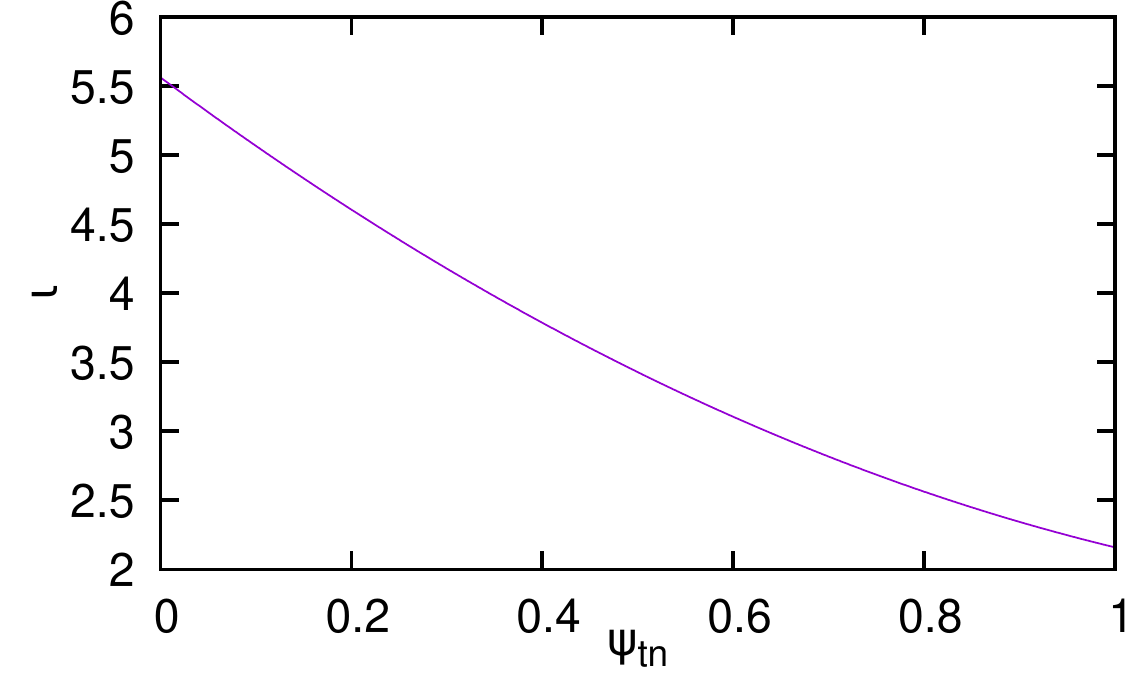}
	\caption{The $\iota$ profile as a function of $\psi_\mathrm{tn}$. This profile is the same for all configurations considered in this paper.}
	\label{fig:iota}
\end{figure}

The consistency checks were done using four equilibria based on the historic Wendelstein 7-A stellarator \cite{w7a1980stabilization} with different values of $\beta$. Note that the $\beta$ values here and throughout the rest of this paper are volume-averaged. These equilibria were intended to be unstable to the (2,1) tearing mode, however, since Wendelstein 7-A had five field periods, the simulations can be done with five-fold periodicity, excluding the unstable $n=1$ Fourier mode and its mode family. Thus, in the computational setting used in this section, there are no physical instabilities. The equilibria were first calculated with NEMEC \cite{strumberger2005user}, and then GVEC was used to refine them. Poloidal modes $m=0,...,12$ and toroidal modes $n=0,5,10,...,50$, which corresponds to $N_\mathrm{ctor} = 21$ and $N_\mathrm{cp} = 5$ in JOREK (see appendix \ref{sec:nagrid}), were used to calculate the equilibrium. All of the equilibria have the same boundary: a rotating ellipse with a minor axis of 0.09131~m and a major axis of 0.1178~m; the major radius of the torus is 1.99~m. The normalized toroidal current profile was also the same for all equilibria:
\begin{equation}
	I_\mathrm{n}(\psi_\mathrm{tn}) = 3\psi_\mathrm{tn} - 3\psi_\mathrm{tn}^2 + \psi_\mathrm{tn}^3,
\end{equation}
where $\psi_\mathrm{tn}$ is the toroidal flux normalized so that $\psi_\mathrm{tn} = 0$ at the axis and $\psi_\mathrm{tn} = 1$ at the boundary. $I_\mathrm{n}(\psi_\mathrm{tn})$, which represents the toroidal current enclosed by the flux surface $\psi_\mathrm{tn}$, is normalized by the total toroidal current, which was 17.5~kA in the cases considered, such that $I_\mathrm{n}(1) = 1$. The total toroidal magnetic flux through a poloidal plane was 0.08~Wb. The pressures at the axis were 1~Pa, 100~Pa, 500~Pa and 1~kPa, which corresponds to $\beta$-values of ${2.3\cdot 10^{-5}~\%}$, $2.3\cdot 10^{-3}~\%$, 0.011~\% and 0.022~\%, respectively. The pressure profiles are given by
\begin{equation}
	p(\psi_\mathrm{tn}) = p_\mathrm{a} - (p_\mathrm{a} - p_\mathrm{b})\psi_\mathrm{tn},
\end{equation}
where $p$ is the pressure in pascals, $p_\mathrm{a}$ is the pressure at the axis and $p_\mathrm{b}$ is the pressure at the boundary. For the $\beta=2.3\cdot 10^{-5}~\%$ ($p_\mathrm{a} = 1$~Pa) case, $p_\mathrm{b} = 0.01$~Pa, while for the other three cases, $p_\mathrm{b} = 1$~Pa. When finding the initial conditions from the GVEC equilibrium, $N_\mathrm{tor} = 9$ and $N_\mathrm{p} = 5$ (see appendix \ref{sec:nagrid}) was used for the variables, which corresponds to Fourier modes $n=0,5,...,20$. The profile of the rotational transform $\iota$, which was the same for all equilibria considered in this paper, is shown in Figure \ref{fig:iota}.

\begin{figure}
	\centering
	\includegraphics[scale=0.75]{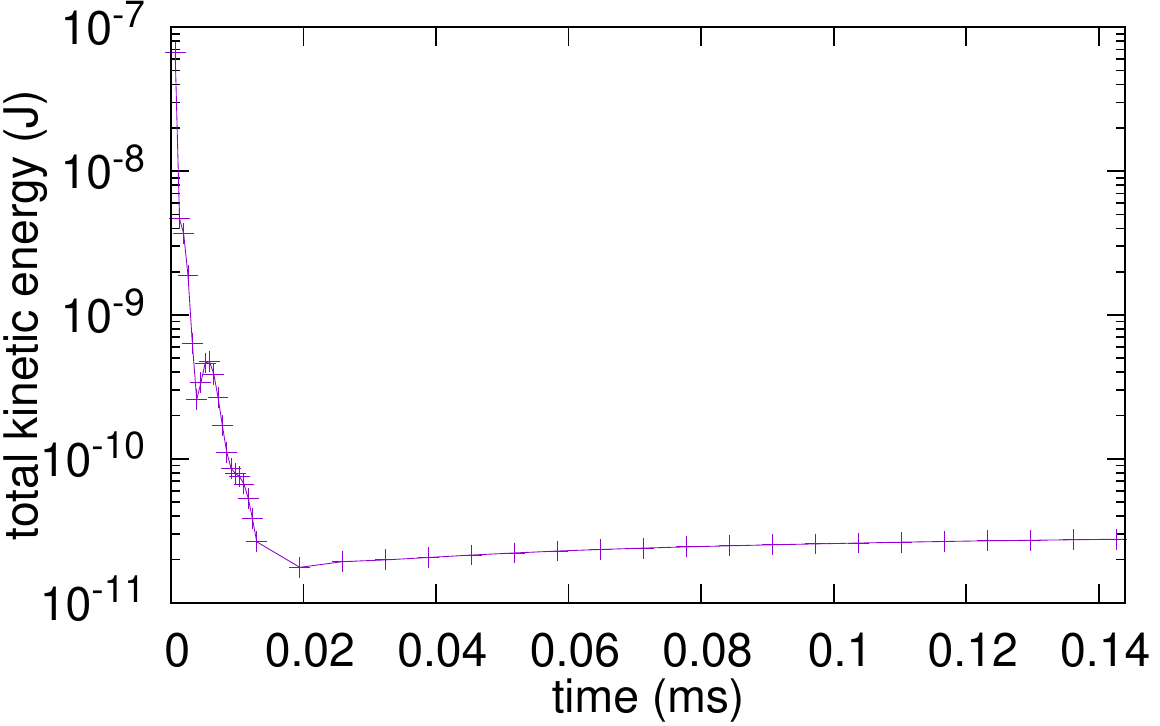}
	\caption{The total kinetic energy of the plasma in the $\beta=0.022~\%$ case during the first 0.144~ms of the simulation (20 time steps of 1 and 20 time steps of 10 JOREK time units) showing the damping out of motion due to the neglect of fourth order terms in the Lorentz force.}
	\label{fig:kinen}
\end{figure}

All of the simulations were run with a spatially constant resistivity $\eta = 1.938\cdot 10^{-9}~\Omega\mathrm{\cdot m}$ and viscosity $\mu = 2.90\cdot 10^{-9}~\mathrm{kg/(m\cdot s)}$. In addition, a hyperviscosity $\mu_h = 2.90\cdot 10^{-12}~\mathrm{kg\cdot m/s}$ was applied. The radial resolution of the finite elements was 41 nodes, and the poloidal resolution was 48 nodes, i.e. moving from the axis to the edge, 41 grid nodes will be passed, counting the nodes at the axis and edge, and 48 grid nodes will be passed in one poloidal turn. The first part of the simulation was run using the implicit Euler time stepping scheme to damp out small oscillations that were present due to the neglect of fourth order terms in the Lorentz force and different discrete representations in JOREK and GVEC (see Figure \ref{fig:kinen}). This consisted of 20 time steps of length $6.484\cdot 10^{-4}~\mathrm{ms}$ (1 in JOREK units), followed by 20 time steps of length $6.484\cdot 10^{-3}~\mathrm{ms}$ (10 in JOREK units), followed by 10 time steps of length $6.484\cdot 10^{-2}~\mathrm{ms}$ (100 in JOREK units). For the $\beta = 0.022\%$ case, but not for the others, this was followed by another 10 time steps of length $6.484\cdot 10^{-2}~\mathrm{ms}$. In the second part of the simulation, the Crank-Nicolson time stepping scheme was used\footnote{Time stepping in JOREK involves first linearizing the equations in time around the current time step, and then applying applying an implicit time advance method, such as the Crank-Nicolson scheme. For more details see Ref \onlinecite{nikulsin2021testing}.}, and all four cases were simulated for 6.484~ms (10000 in JOREK units). The $\beta=2.3\cdot 10^{-5}~\%$ and $\beta=2.3\cdot 10^{-3}~\%$ cases used time steps of length $6.484\cdot 10^{-2}~\mathrm{ms}$ in the second part, however the $\beta=0.011~\%$ and $\beta=0.022~\%$ required shorter time steps ($3.242\cdot 10^{-2}~\mathrm{ms}$ and $1.621\cdot 10^{-2}~\mathrm{ms}$, respectively) for numerical stability. When evaluating the integrals in the weak form of the equations \eqref{eq:rmhd}, the toroidal integration was done by summing over 40 poloidal planes spread evenly over one period.

\begin{figure}
	\centering
	\includegraphics[scale=0.75]{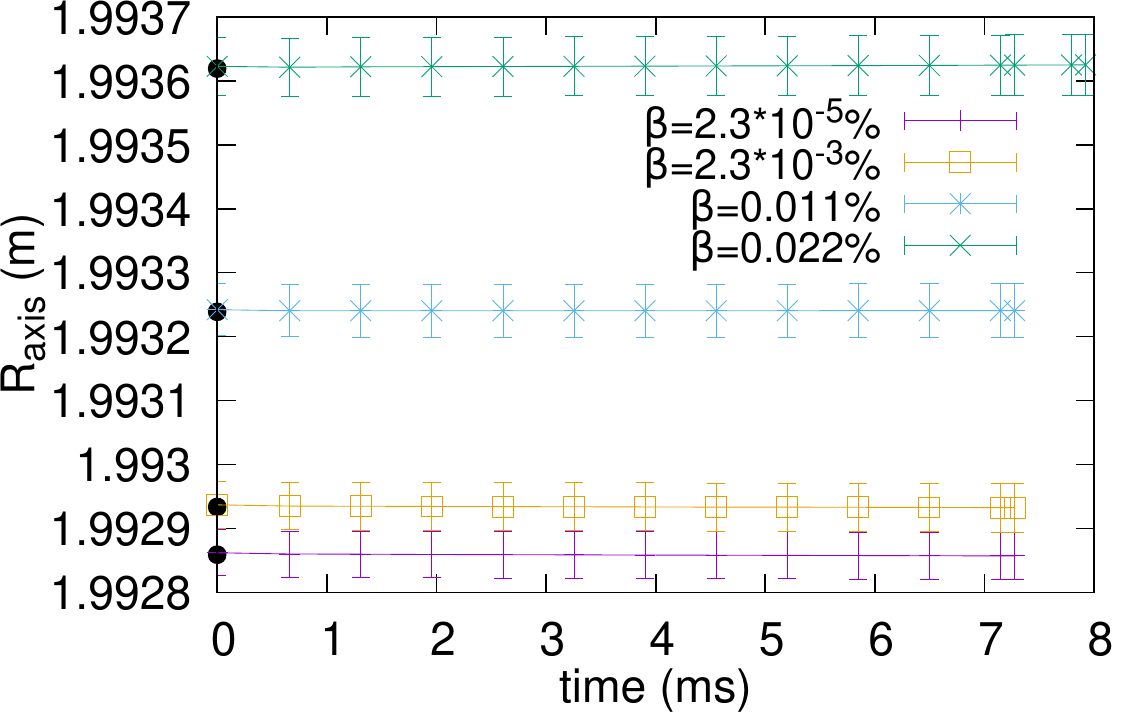}
	\caption{The $R$ coordinate of the magnetic axis as a function of time for the four different $\beta$ cases. The axis position in the full MHD equilibrium as calculated in GVEC is shown by the black dot at $t=0$.}
	\label{fig:axis}
\end{figure}

As expected, no large scale motion was observed in any of the four simulations, confirming in fact that equilibrium is maintained in the reduced MHD model and its implementation. This can be seen in Figure \ref{fig:axis}, where the $R$ coordinate of the magnetic axis is plotted as a function of time for each of the four simulations, along with the error bars. The straight line of the same color outside the error bars represents the axis position in the full MHD equilibrium as calculated in GVEC. The difference between the JOREK and GVEC axis positions is due to the magnetic field being approximated by the reduced MHD ansatz. The axis was determined by making an initial guess for its $(R,z)$ position in the $\phi=0$ poloidal plane, and then tracing the field line at that position for ten toroidal turns, after which the tolerance $T = 0.1\sqrt{(\max R_i - \min R_i)(\max z_i - \min z_i)}$, where $i=1,...,10$, is calculated. If this tolerance is smaller than the cutoff, which was set to $5\cdot 10^{-5}~\mathrm{m}$, then the axis is considered found: the axis position at $\phi=0$ is $(R_\mathrm{c},z_\mathrm{c}) = ((\max R_i + \min R_i)/2,(\max z_i + \min z_i)/2)$. If not, then the field line tracing is restarted at $(R_\mathrm{c},z_\mathrm{c})$, and the process is repeated until the tolerance is less than the cutoff. The error in the $R$-coordinate was was estimated as $E_R = 0.1(\max R_i - \min R_i)$ and plotted as error bars in Figure \ref{fig:axis}.

\begin{figure*}
	\centering
	\includegraphics[scale=0.9]{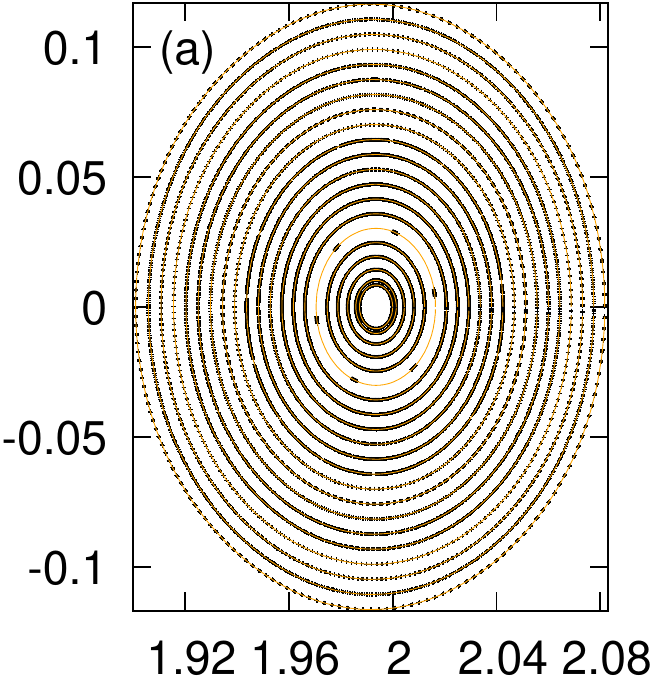}
	\includegraphics[scale=0.9]{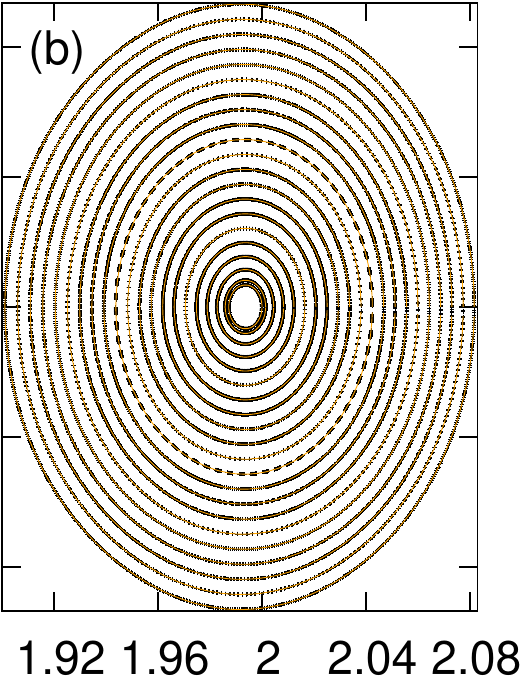}
	\includegraphics[scale=0.9]{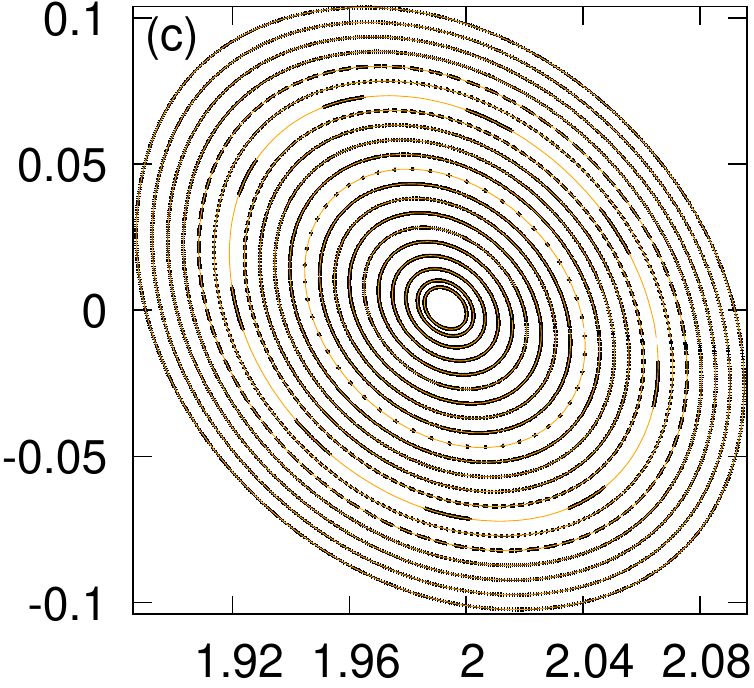}
	\caption{The Poincare plots (black points) for the $\beta=0.022~\%$ case at $t=0$ and $\phi=0$ (a), at $\phi=0$ after the simulation is over ($t=7.275~\mathrm{ms}$) (b), and at $t=0$ and $\phi = 3\pi/10$ (3/4 of the way through one period) (c). The flux surfaces in the full MHD equilibrium as calculated in GVEC are shown by the orange lines.}
	\label{fig:poinc}
\end{figure*}

To demonstrate that there is no significant motion even away from the axis, the Poincare plots for the $\beta=0.022~\%$ case are shown in Figure \ref{fig:poinc}, both before and after the simulation, along with the flux surfaces of the GVEC equilibrium. As can be seen, the flux surfaces in JOREK coincide with the GVEC flux surfaces, so the error introduced by using the reduced MHD ansatz for the magnetic field has no noticeable effect on the flux surfaces. Moreover, the flux surfaces do not move during the simulation, preserving the stable equilibrium as expected.

\section{Tearing mode benchmark}\label{sec:tearing}

\begin{figure}[b]
	\centering
	\includegraphics[scale=1.25]{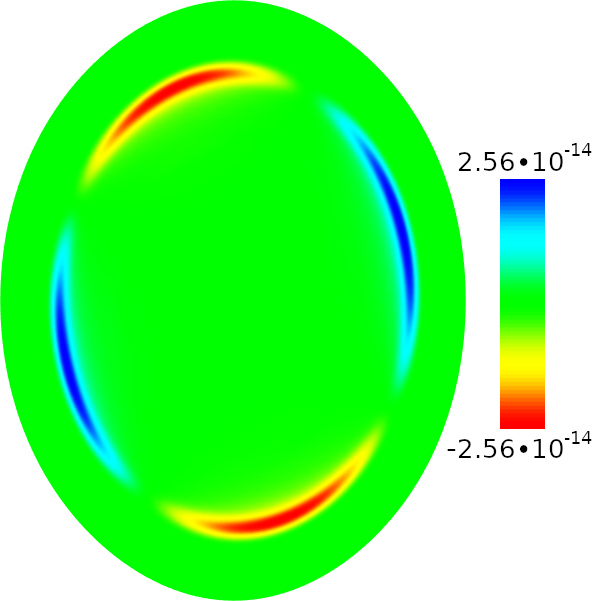}
	\caption{The velocity stream function $\Phi$ without the $n=0,5,10$ Fourier modes on the $\phi = 0$ poloidal plane in the $\beta=2.3\cdot 10^{-5}~\%$, $\eta = 1.938\cdot 10^{-6}~\Omega\mathrm{\cdot m}$ case at $t = 0.969$~ms.}
	\label{fig:tearing}
\end{figure}

Having demonstrated that basic stellarator simulations can be run with the correct equilibrium in the newly implemented model in JOREK, the next step is to simulate instabilities and benchmark them against known results. Tearing modes in the Wendelstein 7-A stellarator were used for this purpose. Three cases at different values of $\beta$ were considered: $2.3\cdot 10^{-5}~\%$, $2.3\cdot 10^{-4}~\%$ and $2.3\cdot 10^{-3}~\%$. For reference, Figure \ref{fig:tearing} shows the velocity stream function $\Phi$ without the $n=0,5,10$ Fourier modes, which do not contribute to the tearing mode, on the $\phi = 0$ poloidal plane during the pre-saturation (linear) phase of the full torus simulation (see below). The characteristic (2,1) structure of the mode is clearly visible. The $\beta=2.3\cdot 10^{-5}~\%$ and $\beta=2.3\cdot 10^{-3}~\%$ are the same equilibria that were used in the previous section, with the $\beta=2.3\cdot 10^{-4}~\%$ being a new equilibrium with the same boundary and current profile as the other two and an intermediate value of $\beta$. In this new intermediate equilibrium, $p_\mathrm{a} = 10$~Pa and $p_\mathrm{b} = 0.1$~Pa. When finding the initial conditions from the GVEC equilibrium, $N_\mathrm{tor} = 5$ and $N_\mathrm{p} = 5$ (see appendix \ref{sec:nagrid}) was used for the variables, which corresponds to Fourier modes $n=0,5,10$.

Just as before, the stellarator simulations were run with the implicit Euler time stepping scheme and five-fold periodicity to damp out oscillations. This consisted of 20 time steps of length $6.484\cdot 10^{-4}~\mathrm{ms}$, followed by 20 time steps of length $6.484\cdot 10^{-3}~\mathrm{ms}$, followed by 5 time steps of length $6.484\cdot 10^{-2}~\mathrm{ms}$. The resistivity was set to $\eta = 1.938\cdot 10^{-6}~\Omega\mathrm{\cdot m}$ and the viscosity was zero. The hyperviscosity was $\mu_h = 2.90\cdot 10^{-15}~\mathrm{kg\cdot m/s}$ for the $\beta=2.3\cdot 10^{-5}~\%$ and the $\beta=2.3\cdot 10^{-4}~\%$ cases and $\mu_h = 7.25\cdot 10^{-15}~\mathrm{kg\cdot m/s}$ for the $\beta=2.3\cdot 10^{-3}~\%$ case. The finite element resolution was 41 nodes radially and 48 nodes poloidally, just as before. Both heat conduction and mass diffusion were set to zero. It should be noted that, when using $N_\mathrm{tor} = 5$, anisotropic transport cannot be properly modelled, as field lines tend to slightly drift from flux surfaces after many toroidal turns, which leads to parallel transport contributing to perpendicular transport after enough time steps. This problem can be remedied by including more toroidal modes.

\begin{figure}
	\centering
	\includegraphics[scale=0.6]{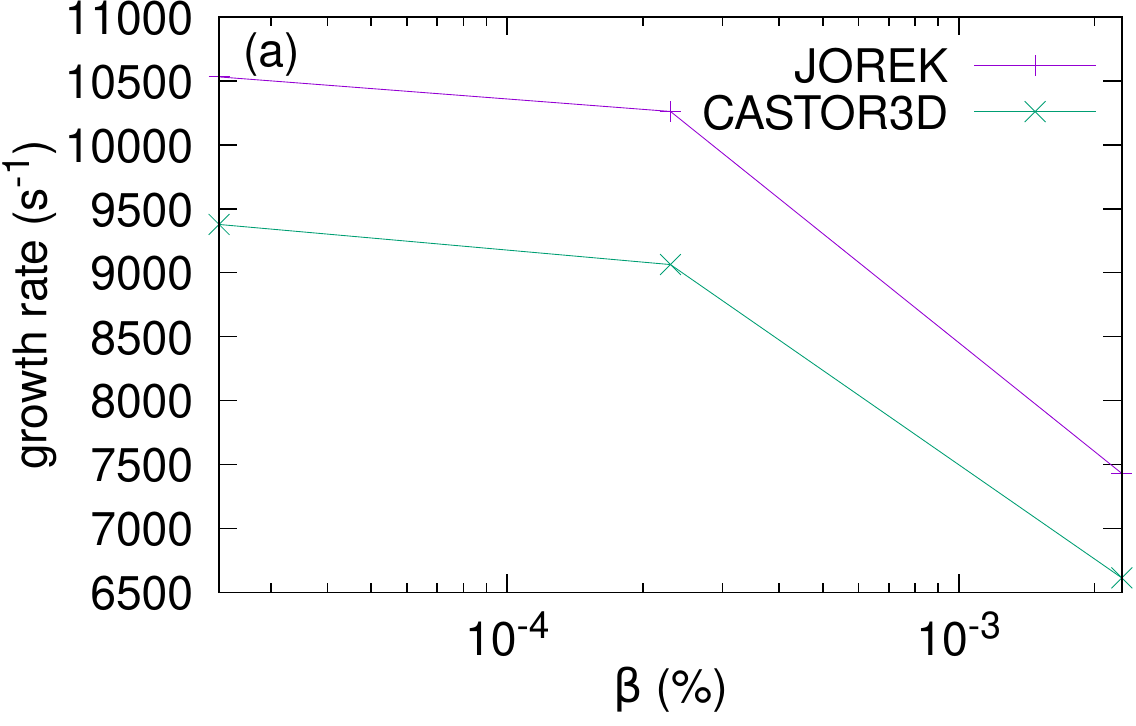}
	\includegraphics[scale=0.6]{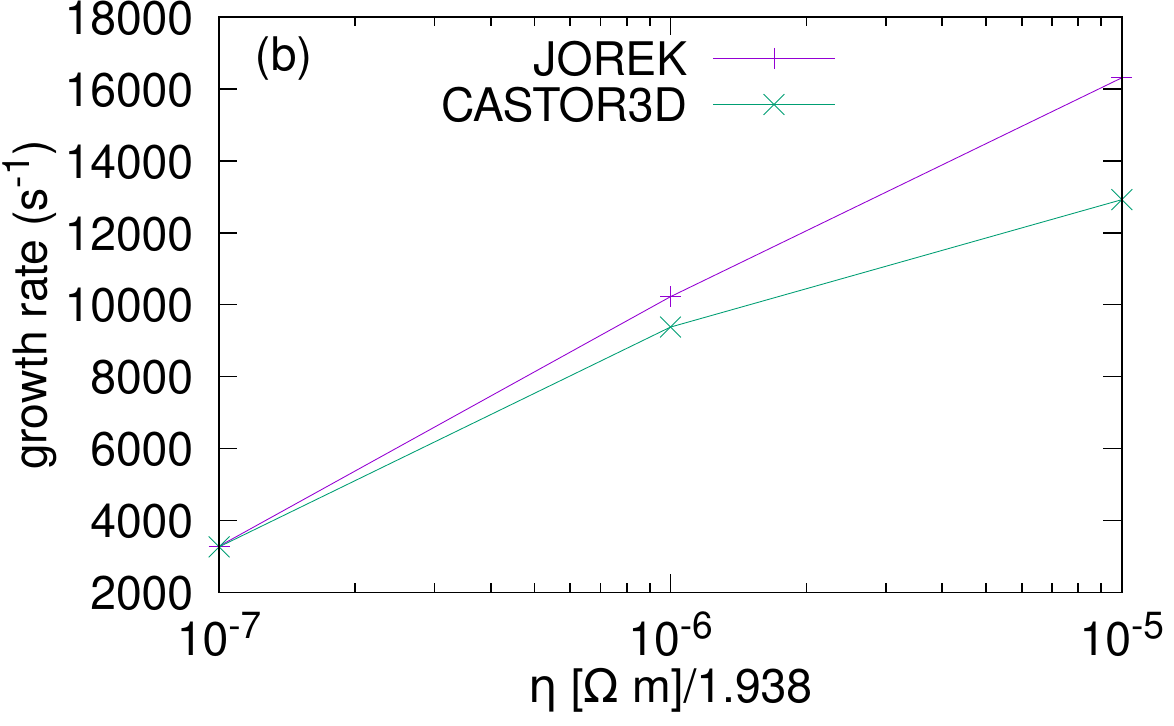}
	\caption{The JOREK and CASTOR3D growth rates at $\eta = 1.938\cdot 10^{-6}~\Omega\mathrm{\cdot m}$ and differing betas (a), and at $\beta=2.3\cdot 10^{-5}~\%$ and differing resistivities (b).}
	\label{fig:grates}
\end{figure}

In the second part of the simulation, the domain was extended to the full torus, taking now into account all of the $n=0,...,10$ Fourier modes, corresponding to $N_\mathrm{tor} = 21$ and $N_\mathrm{p} = 1$ (see appendix \ref{sec:nagrid}). The Crank-Nicolson scheme was used with time steps of length $1.621\cdot 10^{-2}~\mathrm{ms}$. The toroidal integration in the weak form of equations \eqref{eq:rmhd} was done by summing over 40 poloidal planes spread evenly throughout the full torus. The number of Fourier modes, number of poloidal planes and the values of hyperviscosity, resolution and time step size were chosen after scanning over several values for each parameter and choosing the value at which the growth rate of the tearing mode converged. For the present purposes, convergence is considered to be achieved when halving the time step size or hyperviscosity, or doubling the resolution, number of modes or number of planes leads to a change in the growth rate of less than 1.5\%. The convergence test was done for the $\beta=2.3\cdot 10^{-3}~\%$ and $\beta=2.3\cdot 10^{-5}~\%$ cases, resulting in all of the parameters converging to the same values, except for hyperviscosity, which converged to $\mu_h = 7.25\cdot 10^{-15}~\mathrm{kg\cdot m/s}$ for the $\beta=2.3\cdot 10^{-3}~\%$ case and $\mu_h = 2.90\cdot 10^{-15}~\mathrm{kg\cdot m/s}$ for the $\beta=2.3\cdot 10^{-5}~\%$ case. The $\beta=2.3\cdot 10^{-4}~\%$ case was then run using the lower value of hyperviscosity. Figure \ref{fig:grates}~a shows the values of the growth rates from JOREK alongside the values calculated in the linear MHD code CASTOR3D \cite{strumberger2016castor3d,strumberger2019linear}. The maximum deviation between the two codes is 13.2\%, and occurs at $\beta=2.3\cdot 10^{-4}~\%$.

\begin{figure}[b]
	\centering
	\includegraphics[scale=0.6]{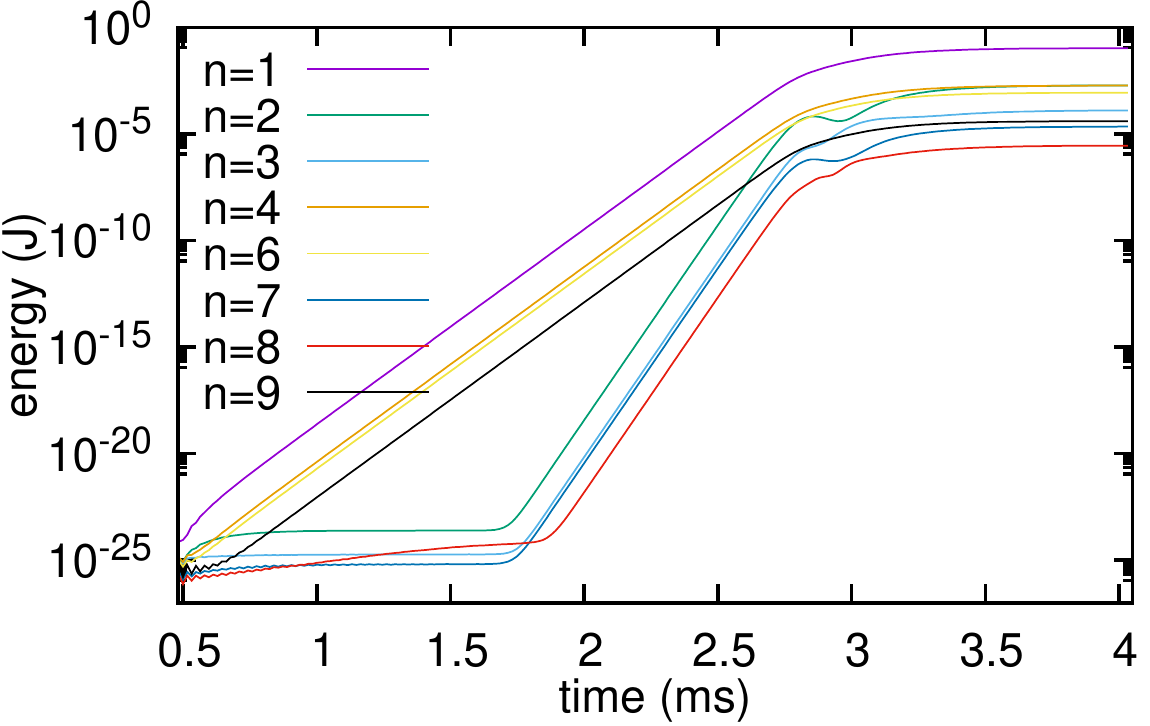}
	\caption{The individual magnetic energies of the Fourier modes $n=1,...,4,6,...,9$ in the $\beta=2.3\cdot 10^{-5}~\%$, $\eta = 1.938\cdot 10^{-6}~\Omega\mathrm{\cdot m}$ case.}
	\label{fig:enmodes}
\end{figure}

\begin{figure}
	\centering
	\includegraphics[scale=0.9]{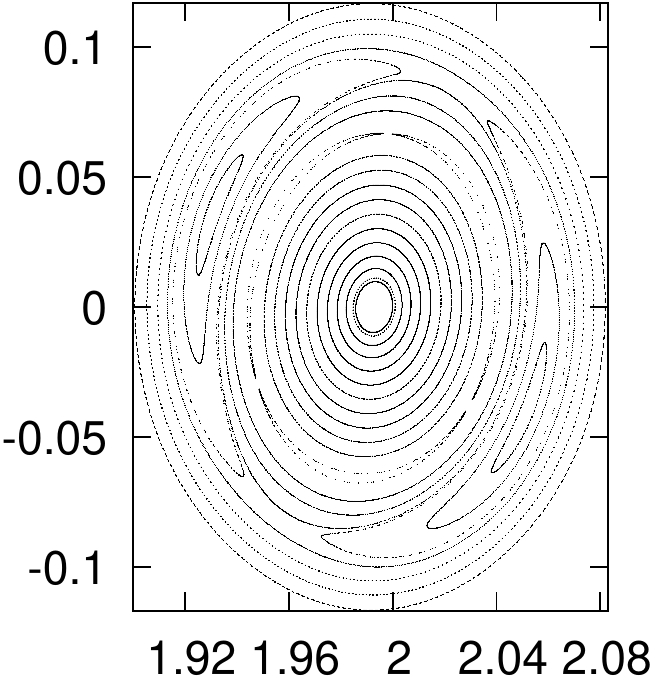}
	\caption{The saturated magnetic island structure in the $\beta=2.3\cdot 10^{-5}~\%$, $\eta = 1.938\cdot 10^{-6}~\Omega\mathrm{\cdot m}$ case at $t = 4.03$~ms.}
	\label{fig:island}
\end{figure}

For reference, Figure \ref{fig:enmodes} shows the magnetic energies of each individual Fourier mode in the $\beta=2.3\cdot 10^{-5}~\%$ case, except for modes that belong to the $n=0$ mode family ($n=0,5,10$). The time axis starts slightly before 0.5~ms, as that is where the full torus simulation begins and the Fourier modes shown are initialized. As expected, the $n=1$ Fourier mode drives the instability, with the rest of its mode family ($n=4,6,9$) growing with it due to linear mode coupling. Around 2~ms, the $n=1$ mode begins to drive the $n=2$ mode via nonlinear coupling, which in turn drives the rest of its mode family ($n=3,7,8$) via linear coupling. Finally, the mode saturates around 3~-~3.5~ms. The saturated magnetic island structure is shown in the Poincare plot in Figure \ref{fig:island}. Note that, aside from the dominant (2,1) island chain, there is a secondary (3,2) island chain towards the interior of the plasma, which is nonlinearly excited by the mode.

Two more simulations were done with the $\beta={2.3\cdot 10^{-5}~\%}$ case, this time using resistivities of $\eta = 1.938\cdot 10^{-7}~\Omega\mathrm{\cdot m}$ and $\eta = 1.938\cdot 10^{-5}~\Omega\mathrm{\cdot m}$, while all of the other parameters were kept the same as before. For the $\eta = 1.938\cdot 10^{-7}~\Omega\mathrm{\cdot m}$ case, a hyperresistivity of $\eta_h = 9.691\cdot 10^{-14}~\Omega\mathrm{\cdot m^2}$ ($5\cdot 10^{-14}$ in JOREK units) had to be introduced in order for the iterative solver to converge in a reasonable amount of time. However, it was first confirmed that introducing this amount of hyperresistivity in the $\eta = 1.938\cdot 10^{-6}~\Omega\mathrm{\cdot m}$ case, which could be run with or without hyperresistivity, changes the growth rate by less than 1.5\%. Figure \ref{fig:grates}~b shows the growth rates for the $\beta=2.3\cdot 10^{-5}~\%$ case at different values of resistivity alongside the growth rates calculated by CASTOR3D. The maximum deviation between the two codes is 26.2\%, occuring at $\eta = 1.938\cdot 10^{-5}~\Omega\mathrm{\cdot m}$. This is most likely due to the neglect of $\vpar$ by the model used in these simulations, as $\vpar$ can be large within the resistive layer, and the size of the resistive layer increases with resistivity. A similar effect is known to exist for quasi-interchange modes. Neglecting the parallel velocity leads to overestimation of the quasi-interchange growth rates by a factor of $\sqrt{1+2q_s^2}$, where $q_s$ is the safety factor of the flux surface where the mode appears \cite{waelbroeck1989nonlinear}. In general, the agreement on the growth rates for the (2,1) tearing mode looks convincing, with deviations on the order of 10\% from CASTOR3D, which solves the linearized full MHD equations.

\section{Ballooning mode benchmark}\label{sec:ballooning}

\begin{figure}[b]
	\centering
	\includegraphics[scale=1.25]{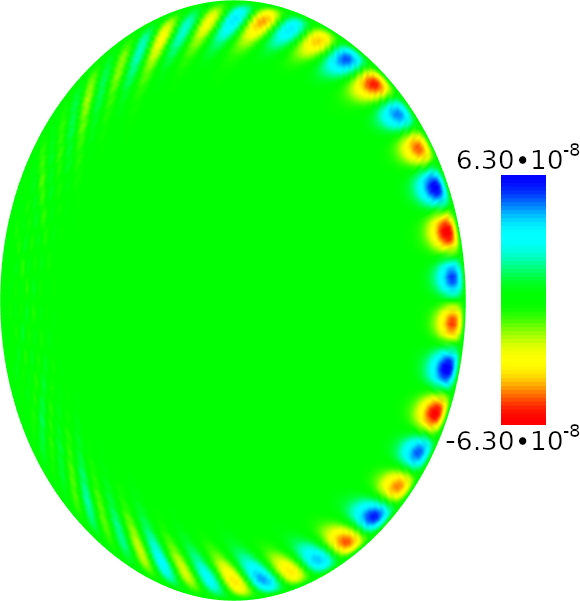}
	\caption{The velocity stream function $\Phi$ on the $\phi = 0$ poloidal plane in the $\beta = 0.21\%$, $\eta = 1.938\cdot 10^{-7}~\Omega\mathrm{\cdot m}$ case at $t = 0.074$~ms.}
	\label{fig:ballooning}
\end{figure}

\begin{figure}
	\centering
	\includegraphics[scale=0.6]{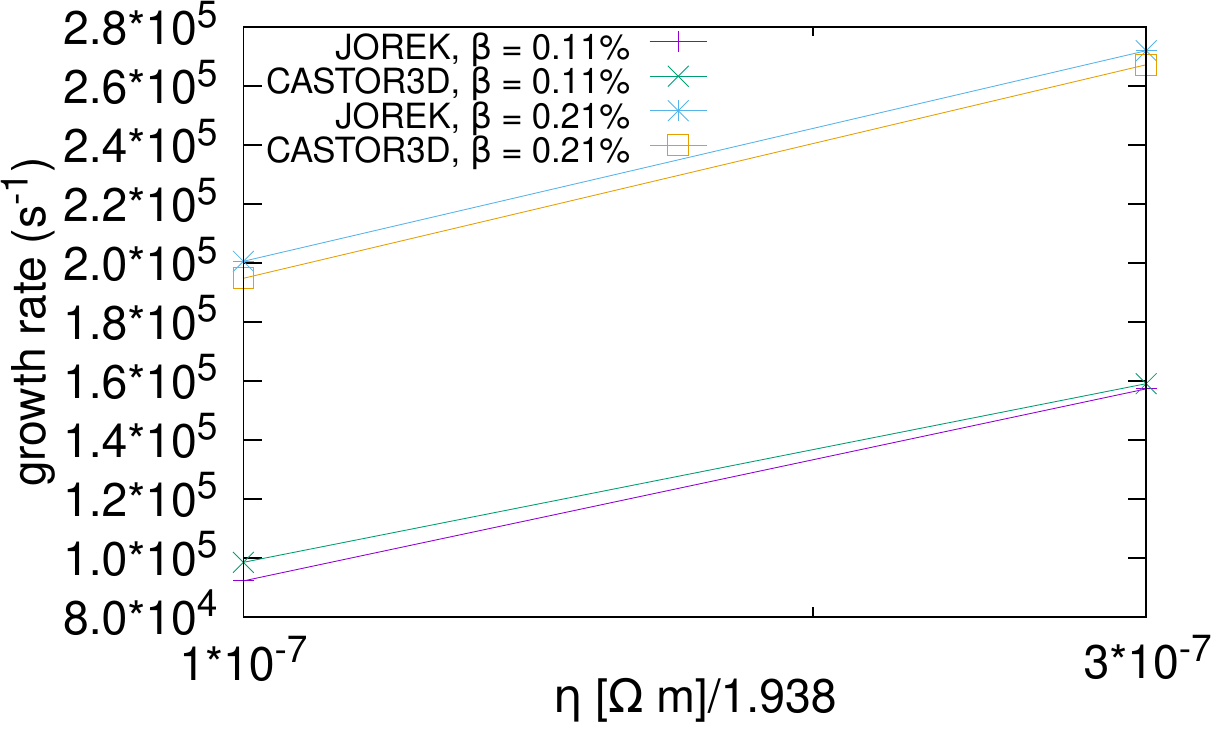}
	\caption{The ballooning mode growth rates as computed by JOREK and CASTOR3D at two different values of $\beta$ and resistivity.}
	\label{fig:bgrates}
\end{figure}

A similar benchmark with CASTOR3D for ballooning mode growth rates in Wendelstein 7-A using equilibria with $\beta = 0.11\%$ and $\beta = 0.21\%$ (corresponding to axis pressures of 5~kPa and 10~kPa) has been done at resistivities of $\eta = 1.938\cdot 10^{-7}~\Omega\mathrm{\cdot m}$ and $\eta = 5.814\cdot 10^{-7}~\Omega\mathrm{\cdot m}$. For reference, the velocity stream function $\Phi$ is shown in Figure \ref{fig:ballooning} during the linear phase of the $\beta = 0.21\%$, $\eta = 1.938\cdot 10^{-7}~\Omega\mathrm{\cdot m}$ simulation. These equilibria have the same toroidal current and pressure profiles as the equilibria in section \ref{sec:cnsis}, with $p_\mathrm{b}$ set to 1~Pa and 100~Pa, respectively. As before, the simulations were initially run with the implicit Euler scheme to damp out oscillations. This first phase of the simulation consisted of 30 time steps of length $6.484\cdot 10^{-4}~\mathrm{ms}$. In the second part of the simulation, the Crank-Nicolson scheme was used, however both parts were run with a five-fold periodicity, since ballooning modes can be simulated with just one period.

Based on linear ballooning mode theory in the tokamak limit, ballooning mode growth rates are known to diverge as $n\to\infty$ if there are no background flows present in the initial equilibrium \cite{freidberg2014ideal}. In order to realistically model ballooning modes, one would have to include an equilibrium flow, such as diamagnetic drift, which will stabilize modes with $n > n_\mathrm{max}$ \cite{huysmans2001modeling}. However, since we only want to do a benchmark, we do not use equilibrium flows in either JOREK or CASTOR3D, but simply cut off the number of Fourier modes in the JOREK simulations, and then limit the CASTOR3D run to the same number of modes. We chose to keep only the $n=0,5,10$ modes in the simulations considered here; the highest mode ($n=10$) is dominant in this case, as it grows the fastest and its energy quickly exceeds that of the other modes. We have also tried including the $n=15$ and $n=20$ modes in case the prediction about increasing growth rates from the tokamak limit is no longer valid for Wendelstein 7-A, however we found that the highest-$n$ mode is the fastest growing one in those simulations as well.

Holding the number of modes fixed at $N_\mathrm{tor} = 5$, a convergence test was done for the $\beta = 0.21\%$ equilibrium while using a resistivity of $\eta = 1.938\cdot 10^{-7}~\Omega\mathrm{\cdot m}$. The growth rate converged at a finite element resolution of 61 nodes radially and 72 nodes poloidally, time step size of $6.484\cdot 10^{-4}~\mathrm{ms}$ and a hyperviscosity of $\mu_h = 7.25\cdot 10^{-15}~\mathrm{kg\cdot m/s}$. The other three simulations were then run with these parameters. Both heat conductivity and mass diffusion parameters were set to zero in all four simulations. Figure \ref{fig:bgrates} shows the growth rates measured in the four cases, along with the growth rates calculated in CASTOR3D for the same four cases. The maximum deviation is 6.4\%, and occurs at $\beta = 0.11\%$, $\eta = 1.938\cdot 10^{-7}~\Omega\cdot\mathrm{m}$, with the other three deviations all being less than 3\%.

Cutting off the Fourier series without stabilizing the high-$n$ modes could lead to spectral blocking\cite{boyd2001chebyshev}, where existing modes couple to higher modes that are not included in the Fourier series, and these coupling contributions are aliased back onto the existing modes, leading to inaccurate results. To check if our results are affected by this numerical error, we first repeated the CASTOR3D run for the $\beta = 0.21\%$, $\eta = 1.938\cdot 10^{-7}~\Omega\cdot\mathrm{m}$ case with the $n=15$ mode included and compared the individual growth rate of the $n=10$ Fourier mode in this simulation to that in the simulation without the $n=15$ mode. We found that the $n=10$ growth rate changed by less than 0.003\%, indicating that the CASTOR3D result is trustworthy. We then took a time step from the $n=0,5,10$ JOREK simulation of the $\beta = 0.21\%$, $\eta = 1.938\cdot 10^{-7}~\Omega\cdot\mathrm{m}$ case where the ballooning mode had already emerged and was growing linearly, and restarted it with the $n=15$ mode also included. We found that the $n=10$ growth rate decreased by about 5-6\% (depending on the exact time step at which it was restarted), becoming much closer to the CASTOR3D value. Thus, in JOREK simulations spectral blocking has more of an effect and it accounts for most of discrepancy between JOREK and CASTOR3D results. Generally, in both codes, the growth rates of the n=10 dominated mode are affected only weakly by the choice of including or excluding the n=15 mode in the simulation.

\section{Conclusion}

Continuing the work of previous papers \cite{nikulsin2019a,nikulsin2021testing}, we implement a stellarator-capable reduced MHD model in JOREK and run several test cases based on the simple geometry of the Wendelstein 7-A stellarator. This paper presents the results, starting with section \ref{sec:rmhd}, which shows that the reduced model introduces an error into Lorentz force, but the error is negligible. The implementation is discussed in sections \ref{sec:domm} and \ref{sec:init}. In order to guarantee $\nabla\cdot\vec B = 0$ to machine precision, an analytical representation of the vacuum magnetic field (i.e. the curl-free component), as derived by Dommaschk \cite{dommaschk1986representations}, was used. This representation is compatible with arbitrary vacuum fields in a toroidal device. In order to run a simulation, the GVEC code is used to calculate an equilibrium, which is then used as an initial condition for the JOREK run. The actual Wendelstein 7-A simulations are presented in sections \ref{sec:cnsis}, \ref{sec:tearing} and \ref{sec:ballooning}. Stable full MHD equilibria are preserved in the reduced model: the flux surfaces do not move throughout the simulation, and closely match the flux surfaces calculated in GVEC, just as one would expect from the ordering argument in section \ref{sec:rmhd}. Further, both tearing and ballooning modes were simulated, and the linear growth rates measured in JOREK are in decent agreement with the growth rates calculated by the CASTOR3D linear full MHD code.

Already in its current form, JOREK is capable of handling more complicated machines, such as Wendelstein 7-X and LHD. Benchmarks involving instabilities in these advanced devices are in progress and the results will be reported in a future publication. We also plan to look for pressure-induced islands in high-$\beta$ simulations see how well their widths match the theoretical predictions of Cary and Kotschenreuther \cite{cary1985pressure}. Studies of scenarios relevant to ongoing experiments are also planned. Of particular interest are the current-driven sawtooth-like crashes observed in Wendelstein 7-X \cite{zanini2020eccd}. Previous studies, which included both linear fully three-dimensional simulations with CASTOR3D \cite{strumberger2020linear} as well as nonlinear simulations in a simplified cylindrical geometry with the TM1 code \cite{yu2020numerical}, have found that the corresponding Wendelstein 7-X equilibria are unstable to single and double tearing modes, as well as resistive kink modes, and that the coupling of double tearing modes with kink modes produces the sawtooth-like crashes. While the family of reduced MHD models used in JOREK, including the models derived in our previous work, cannot accurately reproduce kink modes at higher $\beta$ \cite{hoelzl2020the}, similar sawtooth-like crashes have also been simulated in TM1 at zero $\beta$ \cite{yu2020numerical}. Using JOREK will allow to simulate these modes nonlinearly in a fully three-dimensional geometry.

Another line of work that we intend to pursue is implementing more advanced models for stellarators than the one studied here. The most immediate improvement to the model used here would be to add $\vpar$; other improvements include implementing separate temperatures for electrons and ions, adding a neutral density, and other model extensions already implemented for the tokamak models \cite{hoelzl2020the}. Going further, we also intend to implement a full MHD model for stellarators. This will most likely involve extending the full MHD model described in Ref \onlinecite{haverkort2016implementation}, which uses the standard MHD variables $\{\vec A,\vec v,\rho,T\}$ and doesn't involve any ansatzes or projections, to stellarators. This may require that the vector components of $\vec v$ and $\vec A$ are stored in a flux-surface-aligned coordinate system instead of the cylindrical $(R,z,\phi)$ coordinate system as in Ref \onlinecite{haverkort2016implementation}. However, implementing the full MHD model with the $\{\Psi,\Omega,\Phi,\vpar,\zeta,\rho,T\}$ variables, which was derived in Ref \onlinecite{nikulsin2021testing} and can be seen as a direct extension of the model used in this paper, would also be interesting for comparison purposes. Finally, although we do not expect any issues, it remains to be seen if the model used in this paper will hold up for $\beta\sim 3-5\%$. Further modifications to the model may be needed if future work does not produce satisfactory results for this range of $\beta$.

\section*{Author declarations}

The authors have no conflicts of interest to disclose.

\begin{acknowledgments}
The authors thank Alessandro Zocco and Carolin Nuehrenberg for fruitful discussions, and Michael Drevlak for providing access to his EXTENDER\_P code and patiently answering questions about code use.

This work has been carried out within the framework of the EUROfusion Consortium, funded by the European Union via the Euratom Research and Training Programme (Grant Agreement No 101052200 — EUROfusion). Views and opinions expressed are however those of the author(s) only and do not necessarily reflect those of the European Union or the European Commission. Neither the European Union nor the European Commission can be held responsible for them.
\end{acknowledgments}

\section*{Data availability statement}

The data that supports the findings of this study is available from the corresponding author upon reasonable request.

\appendix
\section{The JOREK non-axisymmetric grid}\label{sec:nagrid}

The spatial discretization in JOREK is done via two-dimensional quadrilateral finite elements in the poloidal plane and a toroidal Fourier expansion. The finite element discretization has $G^1$ continuity, meaning that any discretized functions and their first derivatives are continuous across element boundaries, but second derivatives can jump.

In each element, an element-local coordinate system $(s,t,\phi)$, $s,t\in [0,1]$, is set up, where $(s,t)=(0,0),(0,1),(1,0),(1,1)$ correspond to the four vertices of the element and $\phi$ is the geometric toroidal angle, identical to the $\phi$ coordinate of the cylindrical coordinate system $(R,z,\phi)$. In general, $s$ and $t$ can have arbitrary orientations in the poloidal plane, however in most configurations without an X-point, $s$ is the radial coordinate and $t$ is the poloidal coordinate. All quantities, including the cylindrical coordinates $R$ and $z$, are expressed in terms of the element-local coordinates. Expressing $R$ and $z$ in terms of element-local coordinates allows one to adjust the positions of the vertices of an element, which is normally used to build a flux surface aligned grid. Previously, $R$ and $z$ could only depend on $s$ and $t$, but not $\phi$ \cite{hoelzl2020the}, however this constraint was removed as part of the JOREK stellarator effort. Now, the cylindrical coordinates inside a particular element are represented as:
\begin{equation}
	\{R,z\}(s,t,\phi) = \sum_{n=1}^{N_\mathrm{ctor}} \sum_{i=1}^4 \sum_{j=1}^4 \{R_{ijn},z_{ijn}\} B_{ij}(s,t) Z^\mathrm{c}_n(\phi),
\end{equation}
where $i$ sums over the four vertices of the element, $j$ sums over the degrees of freedom at each vertex and $n$ sums over the toroidal Fourier modes, with $N_\mathrm{ctor}$ being an adjustable parameter. In addition, $B_{ij}(s,t)$ are Bezier basis functions, and
\begin{equation*}
	Z^\mathrm{c}_n(\phi) =
	\begin{cases}
		1, & n = 1 \\
		\cos\left(N_\mathrm{cp}\frac{n}{2}\phi\right), & n~\mathrm{even} \\
		\sin\left(N_\mathrm{cp}\frac{n-1}{2}\phi\right), & n~\mathrm{odd~and}~n>1
	\end{cases}
\end{equation*}
where $N_\mathrm{cp}$ is the periodicity of the underlying geometry. Allowing $R$ and $z$ to depend on $\phi$ makes it possible to build a flux surface aligned grid in a stellarator configuration. The physical quantities, such as density, temperature and $\psi$ are represented in a similar way:
\begin{equation}
	\label{eq:intro:femrep}
	Q(s,t,\phi) = \sum_{n=1}^{N_\mathrm{tor}} \sum_{i=1}^4 \sum_{j=1}^4 Q_{ijn} B_{ij}(s,t) Z_n(\phi).
\end{equation}
Note that $N_\mathrm{tor}$ and $N_\mathrm{ctor}$ are distinct parameters; on a flux surface aligned grid less modes are needed to reasonably represent the physical quantities than the geometry. The Fourier basis function $Z_n(\phi)$ is defined in a similar way to $Z^\mathrm{c}_n(\phi)$, with the difference that $N_\mathrm{cp}$ is replaced by $N_\mathrm{p}$; this allows running full torus simulations without having to add unnecessary modes to the geometry.

\section{Self-adjointness of the linearized reduced MHD operator}\label{sec:linear}

If the equations \eqref{eq:rmhd} are linearized and the non-ideal terms are dropped, then the operator of the resulting linear equation will be self-adjoint. In the ideal case, the equations \eqref{eq:rmhd} can be linearized as follows:
\begin{subequations}
	\begin{gather}
		\nabla\cdot\left(\frac{\rho_0}{B_v^2}\pgrad\tderiv{\Phi}\right) = \nabla\cdot(\widetilde{j}_0\vec B + \widetilde{j}\vec B_0) + B_v\left[\frac{1}{B_v^2},p\right],\label{eq:lphieq} \\
		\tderiv{\rho} = -B_v\left[\frac{\rho_0}{B_v^2},\Phi\right],\label{eq:lrhoeq} \\
		\tderiv{p} = \frac{-1}{B_v}[p_0,\Phi] - \gamma p_0 B_v\left[\frac{1}{B_v^2},\Phi\right],\label{eq:lpeq} \\
		\tderiv{\Psi} = \frac{\llderiv\Phi - [\Psi_0,\Phi]}{B_v},\label{eq:lpsieq}
	\end{gather}
\end{subequations}
where a subscript of 0 denotes the equilibrium value of the corresponding quantity, and perturbations do not have any decorations. Now, analogously to the standard textbook derivation, let $\Phi = \partial\st{\xi}/\partial t$, integrate equations \eqref{eq:lrhoeq}, \eqref{eq:lpeq} and \eqref{eq:lpsieq} over time and insert them back into equation \eqref{eq:lphieq}. The following linear equation is obtained:
\begin{equation}
	\begin{gathered}
		\nabla\cdot\left(\frac{\rho_0}{B_v^2}\pgrad\frac{\partial^2\st{\xi}}{\partial t^2}\right) = L(\st{\xi}), \\
		\begin{aligned}
			L(\st{\xi}) &= \nabla\cdot\Bigg[\widetilde{j}_0\nabla\left(\frac{\llderiv\st{\xi} - [\Psi_0,\st{\xi}]}{B_v}\right)\times\nabla\chi \\
			&+ \vec B_0\Delta^*\left(\frac{\llderiv\st{\xi} - [\Psi_0,\st{\xi}]}{B_v}\right)\Bigg] \\
			&- B_v\left[\frac{1}{B_v^2},\frac{1}{B_v}[p_0,\st{\xi}] + \gamma p_0 B_v\left[\frac{1}{B_v^2},\st{\xi}\right]\right].
		\end{aligned}
	\end{gathered}
\end{equation}
The linear operator $L(\st{\xi})$ is analogous to the force operator $\vec F(\vec\xi)$ of linear full MHD, but has a different dimensionality and cannot be interpreted as the force.

To demonstrate self-adjointness, one has to show that for any two scalar functions $\st{\xi}$ and $\st{\eta}$ which satisfy $\st{\xi} = 0$ and $\st{\eta} = 0$ on the boundary, one has $\int\st{\eta}L(\st{\xi})dV = \int\st{\xi}L(\st{\eta})dV$. Since $L(\st{\xi})$ can be obtained by applying the projection operator \eqref{eq:phipo} to the negative force operator ${-\vec F(\nabla\st{\xi}\times\nabla\chi/B_v^2)}$, one can use the identity $\nabla f\cdot\nabla\times\vec U = -\nabla\cdot(\nabla f\times\vec U)$ and integration by parts to write:
\begin{equation}
	\int\st{\eta}L(\st{\xi})dV = -\int\frac{\nabla\st{\eta}\times\nabla\chi}{B_v^2}\cdot(\vec j_0\times\vec B + \vec j\times\vec B_0 - \nabla p)dV,
\end{equation}
as in Ref \onlinecite{nikulsin2021testing}. Here, $\vec B_0 = \nabla\chi + \nabla\Psi_0\times\nabla\chi$, which allows one to express the magnetic field and pressure perturbations in a familiar way:
\begin{equation}
	\begin{aligned}
		\vec B &= {\nabla\Psi\times\nabla\chi} = \nabla\times\left(\frac{\nabla\st{\xi}\times\nabla\chi}{B_v^2}\times\vec B_0\right), \\
		p &= -[p_0,\st{\xi}]/B_v - \gamma p_0 B_v[B_v^{-2},\st{\xi}] \\
		&= -\nabla p_0\cdot\left(\frac{\nabla\st{\xi}\times\nabla\chi}{B_v^2}\right) - \gamma p_0\nabla\cdot\left(\frac{\nabla\st{\xi}\times\nabla\chi}{B_v^2}\right).
	\end{aligned}
\end{equation}
Now identify $\vec\xi = \nabla\st{\xi}\times\nabla\chi/B_v^2$ and $\vec\eta = \nabla\st{\eta}\times\nabla\chi/B_v^2$, and note that the boundary conditions on $\st{\xi}$ and $\st{\eta}$ imply that the vector fields $\vec\xi$ and $\vec\eta$ satisfy the usual boundary condition for displacement in a plasma surrounded by a wall: $\vec\xi\cdot\vec n = \vec\eta\cdot\vec n = 0$. Finally, one can apply the self-adjointness property of the force operator $\vec F(\vec\xi)$, which allows on to write:
\begin{equation}
	\begin{aligned}
		\int\st{\eta}L(\st{\xi})dV &= -\int\vec\eta\cdot\vec F(\vec\xi)dV = -\int\vec\xi\cdot\vec F(\vec\eta)dV \\
		&= \int\st{\xi}L(\st{\eta})dV.
	\end{aligned}
\end{equation}
This concludes the self-adjointness proof for the linear operator $L(\st{\xi})$.

\bibliography{references}

\begin{thebibliography}{46}%
\makeatletter
\providecommand \@ifxundefined [1]{%
 \@ifx{#1\undefined}
}%
\providecommand \@ifnum [1]{%
 \ifnum #1\expandafter \@firstoftwo
 \else \expandafter \@secondoftwo
 \fi
}%
\providecommand \@ifx [1]{%
 \ifx #1\expandafter \@firstoftwo
 \else \expandafter \@secondoftwo
 \fi
}%
\providecommand \natexlab [1]{#1}%
\providecommand \enquote  [1]{``#1''}%
\providecommand \bibnamefont  [1]{#1}%
\providecommand \bibfnamefont [1]{#1}%
\providecommand \citenamefont [1]{#1}%
\providecommand \href@noop [0]{\@secondoftwo}%
\providecommand \href [0]{\begingroup \@sanitize@url \@href}%
\providecommand \@href[1]{\@@startlink{#1}\@@href}%
\providecommand \@@href[1]{\endgroup#1\@@endlink}%
\providecommand \@sanitize@url [0]{\catcode `\\12\catcode `\$12\catcode
  `\&12\catcode `\#12\catcode `\^12\catcode `\_12\catcode `\%12\relax}%
\providecommand \@@startlink[1]{}%
\providecommand \@@endlink[0]{}%
\providecommand \url  [0]{\begingroup\@sanitize@url \@url }%
\providecommand \@url [1]{\endgroup\@href {#1}{\urlprefix }}%
\providecommand \urlprefix  [0]{URL }%
\providecommand \Eprint [0]{\href }%
\providecommand \doibase [0]{http://dx.doi.org/}%
\providecommand \selectlanguage [0]{\@gobble}%
\providecommand \bibinfo  [0]{\@secondoftwo}%
\providecommand \bibfield  [0]{\@secondoftwo}%
\providecommand \translation [1]{[#1]}%
\providecommand \BibitemOpen [0]{}%
\providecommand \bibitemStop [0]{}%
\providecommand \bibitemNoStop [0]{.\EOS\space}%
\providecommand \EOS [0]{\spacefactor3000\relax}%
\providecommand \BibitemShut  [1]{\csname bibitem#1\endcsname}%
\let\auto@bib@innerbib\@empty
\bibitem [{\citenamefont {Hoelzl}\ \emph {et~al.}(2021)\citenamefont {Hoelzl},
  \citenamefont {Huijsmans}, \citenamefont {Pamela}, \citenamefont
  {B{\'{e}}coulet}, \citenamefont {Nardon}, \citenamefont {Artola},
  \citenamefont {Nkonga}, \citenamefont {Atanasiu}, \citenamefont {Bandaru},
  \citenamefont {Bhole}, \citenamefont {Bonfiglio}, \citenamefont {Cathey},
  \citenamefont {Czarny}, \citenamefont {Dvornova}, \citenamefont
  {Feh{\'{e}}r}, \citenamefont {Fil}, \citenamefont {Franck}, \citenamefont
  {Futatani}, \citenamefont {Gruca}, \citenamefont {Guillard}, \citenamefont
  {Haverkort}, \citenamefont {Holod}, \citenamefont {Hu}, \citenamefont {Kim},
  \citenamefont {Korving}, \citenamefont {Kos}, \citenamefont {Krebs},
  \citenamefont {Kripner}, \citenamefont {Latu}, \citenamefont {Liu},
  \citenamefont {Merkel}, \citenamefont {Meshcheriakov}, \citenamefont
  {Mitterauer}, \citenamefont {Mochalskyy}, \citenamefont {Morales},
  \citenamefont {Nies}, \citenamefont {Nikulsin}, \citenamefont {Orain},
  \citenamefont {Pratt}, \citenamefont {Ramasamy}, \citenamefont {Ramet},
  \citenamefont {Reux}, \citenamefont {Särkimäki}, \citenamefont {Schwarz},
  \citenamefont {Verma}, \citenamefont {Smith}, \citenamefont {Sommariva},
  \citenamefont {Strumberger}, \citenamefont {van Vugt}, \citenamefont
  {Verbeek}, \citenamefont {Westerhof}, \citenamefont {Wieschollek},\ and\
  \citenamefont {Zielinski}}]{hoelzl2020the}%
  \BibitemOpen
  \bibfield  {author} {\bibinfo {author} {\bibfnamefont {M.}~\bibnamefont
  {Hoelzl}}, \bibinfo {author} {\bibfnamefont {G.}~\bibnamefont {Huijsmans}},
  \bibinfo {author} {\bibfnamefont {S.}~\bibnamefont {Pamela}}, \bibinfo
  {author} {\bibfnamefont {M.}~\bibnamefont {B{\'{e}}coulet}}, \bibinfo
  {author} {\bibfnamefont {E.}~\bibnamefont {Nardon}}, \bibinfo {author}
  {\bibfnamefont {F.}~\bibnamefont {Artola}}, \bibinfo {author} {\bibfnamefont
  {B.}~\bibnamefont {Nkonga}}, \bibinfo {author} {\bibfnamefont
  {C.}~\bibnamefont {Atanasiu}}, \bibinfo {author} {\bibfnamefont
  {V.}~\bibnamefont {Bandaru}}, \bibinfo {author} {\bibfnamefont
  {A.}~\bibnamefont {Bhole}}, \bibinfo {author} {\bibfnamefont
  {D.}~\bibnamefont {Bonfiglio}}, \bibinfo {author} {\bibfnamefont
  {A.}~\bibnamefont {Cathey}}, \bibinfo {author} {\bibfnamefont
  {O.}~\bibnamefont {Czarny}}, \bibinfo {author} {\bibfnamefont
  {A.}~\bibnamefont {Dvornova}}, \bibinfo {author} {\bibfnamefont
  {T.}~\bibnamefont {Feh{\'{e}}r}}, \bibinfo {author} {\bibfnamefont
  {A.}~\bibnamefont {Fil}}, \bibinfo {author} {\bibfnamefont {E.}~\bibnamefont
  {Franck}}, \bibinfo {author} {\bibfnamefont {S.}~\bibnamefont {Futatani}},
  \bibinfo {author} {\bibfnamefont {M.}~\bibnamefont {Gruca}}, \bibinfo
  {author} {\bibfnamefont {H.}~\bibnamefont {Guillard}}, \bibinfo {author}
  {\bibfnamefont {J.}~\bibnamefont {Haverkort}}, \bibinfo {author}
  {\bibfnamefont {I.}~\bibnamefont {Holod}}, \bibinfo {author} {\bibfnamefont
  {D.}~\bibnamefont {Hu}}, \bibinfo {author} {\bibfnamefont {S.}~\bibnamefont
  {Kim}}, \bibinfo {author} {\bibfnamefont {S.}~\bibnamefont {Korving}},
  \bibinfo {author} {\bibfnamefont {L.}~\bibnamefont {Kos}}, \bibinfo {author}
  {\bibfnamefont {I.}~\bibnamefont {Krebs}}, \bibinfo {author} {\bibfnamefont
  {L.}~\bibnamefont {Kripner}}, \bibinfo {author} {\bibfnamefont
  {G.}~\bibnamefont {Latu}}, \bibinfo {author} {\bibfnamefont {F.}~\bibnamefont
  {Liu}}, \bibinfo {author} {\bibfnamefont {P.}~\bibnamefont {Merkel}},
  \bibinfo {author} {\bibfnamefont {D.}~\bibnamefont {Meshcheriakov}}, \bibinfo
  {author} {\bibfnamefont {V.}~\bibnamefont {Mitterauer}}, \bibinfo {author}
  {\bibfnamefont {S.}~\bibnamefont {Mochalskyy}}, \bibinfo {author}
  {\bibfnamefont {J.}~\bibnamefont {Morales}}, \bibinfo {author} {\bibfnamefont
  {R.}~\bibnamefont {Nies}}, \bibinfo {author} {\bibfnamefont {N.}~\bibnamefont
  {Nikulsin}}, \bibinfo {author} {\bibfnamefont {F.}~\bibnamefont {Orain}},
  \bibinfo {author} {\bibfnamefont {J.}~\bibnamefont {Pratt}}, \bibinfo
  {author} {\bibfnamefont {R.}~\bibnamefont {Ramasamy}}, \bibinfo {author}
  {\bibfnamefont {P.}~\bibnamefont {Ramet}}, \bibinfo {author} {\bibfnamefont
  {C.}~\bibnamefont {Reux}}, \bibinfo {author} {\bibfnamefont {K.}~\bibnamefont
  {Särkimäki}}, \bibinfo {author} {\bibfnamefont {N.}~\bibnamefont
  {Schwarz}}, \bibinfo {author} {\bibfnamefont {P.~S.}\ \bibnamefont {Verma}},
  \bibinfo {author} {\bibfnamefont {S.}~\bibnamefont {Smith}}, \bibinfo
  {author} {\bibfnamefont {C.}~\bibnamefont {Sommariva}}, \bibinfo {author}
  {\bibfnamefont {E.}~\bibnamefont {Strumberger}}, \bibinfo {author}
  {\bibfnamefont {D.}~\bibnamefont {van Vugt}}, \bibinfo {author}
  {\bibfnamefont {M.}~\bibnamefont {Verbeek}}, \bibinfo {author} {\bibfnamefont
  {E.}~\bibnamefont {Westerhof}}, \bibinfo {author} {\bibfnamefont
  {F.}~\bibnamefont {Wieschollek}}, \ and\ \bibinfo {author} {\bibfnamefont
  {J.}~\bibnamefont {Zielinski}},\ }\href {\doibase 10.1088/1741-4326/abf99f}
  {\bibfield  {journal} {\bibinfo  {journal} {Nuclear Fusion}\ }\textbf
  {\bibinfo {volume} {61}},\ \bibinfo {pages} {065001} (\bibinfo {year}
  {2021})}\BibitemShut {NoStop}%
\bibitem [{\citenamefont {Boozer}(2019)}]{boozer2019curl}%
  \BibitemOpen
  \bibfield  {author} {\bibinfo {author} {\bibfnamefont {A.~H.}\ \bibnamefont
  {Boozer}},\ }\href {\doibase 10.1063/1.5116721} {\bibfield  {journal}
  {\bibinfo  {journal} {Physics of Plasmas}\ }\textbf {\bibinfo {volume}
  {26}},\ \bibinfo {pages} {102504} (\bibinfo {year} {2019})}\BibitemShut
  {NoStop}%
\bibitem [{\citenamefont {Helander}(2014)}]{helander2014theory}%
  \BibitemOpen
  \bibfield  {author} {\bibinfo {author} {\bibfnamefont {P.}~\bibnamefont
  {Helander}},\ }\href {\doibase 10.1088/0034-4885/77/8/087001} {\bibfield
  {journal} {\bibinfo  {journal} {Reports on Progress in Physics}\ }\textbf
  {\bibinfo {volume} {77}},\ \bibinfo {pages} {087001} (\bibinfo {year}
  {2014})}\BibitemShut {NoStop}%
\bibitem [{\citenamefont {Freidberg}(2014)}]{freidberg2014ideal}%
  \BibitemOpen
  \bibfield  {author} {\bibinfo {author} {\bibfnamefont {J.~P.}\ \bibnamefont
  {Freidberg}},\ }\href {\doibase 10.1017/CBO9780511795046} {\emph {\bibinfo
  {title} {Ideal MHD}}}\ (\bibinfo  {publisher} {Cambridge University Press},\
  \bibinfo {year} {2014})\BibitemShut {NoStop}%
\bibitem [{\citenamefont {Wakatani}(1978)}]{wakatani1978non}%
  \BibitemOpen
  \bibfield  {author} {\bibinfo {author} {\bibfnamefont {M.}~\bibnamefont
  {Wakatani}},\ }\href {\doibase 10.1088/0029-5515/18/11/003} {\bibfield
  {journal} {\bibinfo  {journal} {Nuclear Fusion}\ }\textbf {\bibinfo {volume}
  {18}},\ \bibinfo {pages} {1499} (\bibinfo {year} {1978})}\BibitemShut
  {NoStop}%
\bibitem [{\citenamefont {Wakatani}, \citenamefont {Shirai},\ and\
  \citenamefont {Yamagiwa}(1984)}]{wakatani1984pressure}%
  \BibitemOpen
  \bibfield  {author} {\bibinfo {author} {\bibfnamefont {M.}~\bibnamefont
  {Wakatani}}, \bibinfo {author} {\bibfnamefont {H.}~\bibnamefont {Shirai}}, \
  and\ \bibinfo {author} {\bibfnamefont {M.}~\bibnamefont {Yamagiwa}},\ }\href
  {\doibase 10.1088/0029-5515/24/11/003} {\bibfield  {journal} {\bibinfo
  {journal} {Nuclear Fusion}\ }\textbf {\bibinfo {volume} {24}},\ \bibinfo
  {pages} {1407} (\bibinfo {year} {1984})}\BibitemShut {NoStop}%
\bibitem [{\citenamefont {Ishii}\ and\ \citenamefont
  {Wakatani}(1995)}]{ishii1995nonlinear}%
  \BibitemOpen
  \bibfield  {author} {\bibinfo {author} {\bibfnamefont {Y.}~\bibnamefont
  {Ishii}}\ and\ \bibinfo {author} {\bibfnamefont {M.}~\bibnamefont
  {Wakatani}},\ }\href {\doibase 10.1088/0741-3335/37/8/003} {\bibfield
  {journal} {\bibinfo  {journal} {Plasma Physics and Controlled Fusion}\
  }\textbf {\bibinfo {volume} {37}},\ \bibinfo {pages} {867} (\bibinfo {year}
  {1995})}\BibitemShut {NoStop}%
\bibitem [{\citenamefont {Strauss}\ \emph {et~al.}(2004)\citenamefont
  {Strauss}, \citenamefont {Sugiyama}, \citenamefont {Fu}, \citenamefont
  {Park},\ and\ \citenamefont {Breslau}}]{strauss2004simulation}%
  \BibitemOpen
  \bibfield  {author} {\bibinfo {author} {\bibfnamefont {H.}~\bibnamefont
  {Strauss}}, \bibinfo {author} {\bibfnamefont {L.}~\bibnamefont {Sugiyama}},
  \bibinfo {author} {\bibfnamefont {G.}~\bibnamefont {Fu}}, \bibinfo {author}
  {\bibfnamefont {W.}~\bibnamefont {Park}}, \ and\ \bibinfo {author}
  {\bibfnamefont {J.}~\bibnamefont {Breslau}},\ }\href {\doibase
  10.1088/0029-5515/44/9/010} {\bibfield  {journal} {\bibinfo  {journal}
  {Nuclear Fusion}\ }\textbf {\bibinfo {volume} {44}},\ \bibinfo {pages} {1008}
  (\bibinfo {year} {2004})}\BibitemShut {NoStop}%
\bibitem [{\citenamefont {Mizuguchi}, \citenamefont {Suzuki},\ and\
  \citenamefont {Ohyabu}(2009)}]{mizuguchi2009nonlinear}%
  \BibitemOpen
  \bibfield  {author} {\bibinfo {author} {\bibfnamefont {N.}~\bibnamefont
  {Mizuguchi}}, \bibinfo {author} {\bibfnamefont {Y.}~\bibnamefont {Suzuki}}, \
  and\ \bibinfo {author} {\bibfnamefont {N.}~\bibnamefont {Ohyabu}},\ }\href
  {\doibase 10.1088/0029-5515/49/9/095023} {\bibfield  {journal} {\bibinfo
  {journal} {Nuclear Fusion}\ }\textbf {\bibinfo {volume} {49}},\ \bibinfo
  {pages} {095023} (\bibinfo {year} {2009})}\BibitemShut {NoStop}%
\bibitem [{\citenamefont {Zhou}\ \emph {et~al.}(2021)\citenamefont {Zhou},
  \citenamefont {Ferraro}, \citenamefont {Jardin},\ and\ \citenamefont
  {Strauss}}]{zhou2021approach}%
  \BibitemOpen
  \bibfield  {author} {\bibinfo {author} {\bibfnamefont {Y.}~\bibnamefont
  {Zhou}}, \bibinfo {author} {\bibfnamefont {N.}~\bibnamefont {Ferraro}},
  \bibinfo {author} {\bibfnamefont {S.}~\bibnamefont {Jardin}}, \ and\ \bibinfo
  {author} {\bibfnamefont {H.}~\bibnamefont {Strauss}},\ }\href {\doibase
  10.1088/1741-4326/ac0b35} {\bibfield  {journal} {\bibinfo  {journal} {Nuclear
  Fusion}\ }\textbf {\bibinfo {volume} {61}},\ \bibinfo {pages} {086015}
  (\bibinfo {year} {2021})}\BibitemShut {NoStop}%
\bibitem [{\citenamefont {Sato}\ \emph {et~al.}(2017)\citenamefont {Sato},
  \citenamefont {Nakajima}, \citenamefont {Watanabe},\ and\ \citenamefont
  {Todo}}]{sato2017characteristics}%
  \BibitemOpen
  \bibfield  {author} {\bibinfo {author} {\bibfnamefont {M.}~\bibnamefont
  {Sato}}, \bibinfo {author} {\bibfnamefont {N.}~\bibnamefont {Nakajima}},
  \bibinfo {author} {\bibfnamefont {K.}~\bibnamefont {Watanabe}}, \ and\
  \bibinfo {author} {\bibfnamefont {Y.}~\bibnamefont {Todo}},\ }\href {\doibase
  10.1088/1741-4326/aa8492} {\bibfield  {journal} {\bibinfo  {journal} {Nuclear
  Fusion}\ }\textbf {\bibinfo {volume} {57}},\ \bibinfo {pages} {126023}
  (\bibinfo {year} {2017})}\BibitemShut {NoStop}%
\bibitem [{\citenamefont {Sovinec}\ \emph {et~al.}(2020)\citenamefont
  {Sovinec}, \citenamefont {Guilbault}, \citenamefont {Cornille},\ and\
  \citenamefont {Bechtel}}]{sovinec2020development}%
  \BibitemOpen
  \bibfield  {author} {\bibinfo {author} {\bibfnamefont {C.~R.}\ \bibnamefont
  {Sovinec}}, \bibinfo {author} {\bibfnamefont {C.~M.}\ \bibnamefont
  {Guilbault}}, \bibinfo {author} {\bibfnamefont {B.~S.}\ \bibnamefont
  {Cornille}}, \ and\ \bibinfo {author} {\bibfnamefont {T.~A.}\ \bibnamefont
  {Bechtel}},\ }\href {https://meetings.aps.org/Meeting/DPP20/Session/BO05.9}
  {\enquote {\bibinfo {title} {Development of {MHD} simulation capability for
  stellarators},}\ }\bibinfo {howpublished} {62nd Annual Meeting of the
  APS-DPP} (\bibinfo {year} {2020})\BibitemShut {NoStop}%
\bibitem [{\citenamefont {Schlutt}\ \emph {et~al.}(2012)\citenamefont
  {Schlutt}, \citenamefont {Hegna}, \citenamefont {Sovinec}, \citenamefont
  {Knowlton},\ and\ \citenamefont {Hebert}}]{schlutt2012numerical}%
  \BibitemOpen
  \bibfield  {author} {\bibinfo {author} {\bibfnamefont {M.}~\bibnamefont
  {Schlutt}}, \bibinfo {author} {\bibfnamefont {C.}~\bibnamefont {Hegna}},
  \bibinfo {author} {\bibfnamefont {C.}~\bibnamefont {Sovinec}}, \bibinfo
  {author} {\bibfnamefont {S.}~\bibnamefont {Knowlton}}, \ and\ \bibinfo
  {author} {\bibfnamefont {J.}~\bibnamefont {Hebert}},\ }\href {\doibase
  10.1088/0029-5515/52/10/103023} {\bibfield  {journal} {\bibinfo  {journal}
  {Nuclear Fusion}\ }\textbf {\bibinfo {volume} {52}},\ \bibinfo {pages}
  {103023} (\bibinfo {year} {2012})}\BibitemShut {NoStop}%
\bibitem [{\citenamefont {Hindenlang}\ \emph {et~al.}(2012)\citenamefont
  {Hindenlang}, \citenamefont {Gassner}, \citenamefont {Altmann}, \citenamefont
  {Beck}, \citenamefont {Staudenmaier},\ and\ \citenamefont
  {Munz}}]{hindenlang2021explicit}%
  \BibitemOpen
  \bibfield  {author} {\bibinfo {author} {\bibfnamefont {F.~J.}\ \bibnamefont
  {Hindenlang}}, \bibinfo {author} {\bibfnamefont {G.~J.}\ \bibnamefont
  {Gassner}}, \bibinfo {author} {\bibfnamefont {C.}~\bibnamefont {Altmann}},
  \bibinfo {author} {\bibfnamefont {A.}~\bibnamefont {Beck}}, \bibinfo {author}
  {\bibfnamefont {M.}~\bibnamefont {Staudenmaier}}, \ and\ \bibinfo {author}
  {\bibfnamefont {C.-D.}\ \bibnamefont {Munz}},\ }\href@noop {} {\bibfield
  {journal} {\bibinfo  {journal} {Computers \& Fluids}\ }\textbf {\bibinfo
  {volume} {61}},\ \bibinfo {pages} {86} (\bibinfo {year} {2012})}\BibitemShut
  {NoStop}%
\bibitem [{\citenamefont {Huysmans}\ and\ \citenamefont
  {Czarny}(2007)}]{huysmans2007mhd}%
  \BibitemOpen
  \bibfield  {author} {\bibinfo {author} {\bibfnamefont {G.}~\bibnamefont
  {Huysmans}}\ and\ \bibinfo {author} {\bibfnamefont {O.}~\bibnamefont
  {Czarny}},\ }\href {\doibase 10.1088/0029-5515/47/7/016} {\bibfield
  {journal} {\bibinfo  {journal} {Nuclear Fusion}\ }\textbf {\bibinfo {volume}
  {47}},\ \bibinfo {pages} {659} (\bibinfo {year} {2007})}\BibitemShut
  {NoStop}%
\bibitem [{\citenamefont {Czarny}\ and\ \citenamefont
  {Huysmans}(2008)}]{czarny2008bezier}%
  \BibitemOpen
  \bibfield  {author} {\bibinfo {author} {\bibfnamefont {O.}~\bibnamefont
  {Czarny}}\ and\ \bibinfo {author} {\bibfnamefont {G.}~\bibnamefont
  {Huysmans}},\ }\href {\doibase 10.1016/j.jcp.2008.04.001} {\bibfield
  {journal} {\bibinfo  {journal} {Journal of Computational Physics}\ }\textbf
  {\bibinfo {volume} {227}},\ \bibinfo {pages} {7423} (\bibinfo {year}
  {2008})}\BibitemShut {NoStop}%
\bibitem [{\citenamefont {Breslau}, \citenamefont {Ferraro},\ and\
  \citenamefont {Jardin}(2009)}]{breslau2009some}%
  \BibitemOpen
  \bibfield  {author} {\bibinfo {author} {\bibfnamefont {J.}~\bibnamefont
  {Breslau}}, \bibinfo {author} {\bibfnamefont {N.}~\bibnamefont {Ferraro}}, \
  and\ \bibinfo {author} {\bibfnamefont {S.}~\bibnamefont {Jardin}},\ }\href
  {\doibase 10.1063/1.3224035} {\bibfield  {journal} {\bibinfo  {journal}
  {Physics of Plasmas}\ }\textbf {\bibinfo {volume} {16}},\ \bibinfo {pages}
  {092503} (\bibinfo {year} {2009})}\BibitemShut {NoStop}%
\bibitem [{\citenamefont {Izzo}\ \emph {et~al.}(1985)\citenamefont {Izzo},
  \citenamefont {Monticello}, \citenamefont {DeLucia}, \citenamefont {Park},\
  and\ \citenamefont {Ryu}}]{izzo1985reduced}%
  \BibitemOpen
  \bibfield  {author} {\bibinfo {author} {\bibfnamefont {R.}~\bibnamefont
  {Izzo}}, \bibinfo {author} {\bibfnamefont {D.}~\bibnamefont {Monticello}},
  \bibinfo {author} {\bibfnamefont {J.}~\bibnamefont {DeLucia}}, \bibinfo
  {author} {\bibfnamefont {W.}~\bibnamefont {Park}}, \ and\ \bibinfo {author}
  {\bibfnamefont {C.}~\bibnamefont {Ryu}},\ }\href {\doibase 10.1063/1.865061}
  {\bibfield  {journal} {\bibinfo  {journal} {The Physics of fluids}\ }\textbf
  {\bibinfo {volume} {28}},\ \bibinfo {pages} {903} (\bibinfo {year}
  {1985})}\BibitemShut {NoStop}%
\bibitem [{\citenamefont {Strauss}(1997)}]{strauss1997reduced}%
  \BibitemOpen
  \bibfield  {author} {\bibinfo {author} {\bibfnamefont {H.}~\bibnamefont
  {Strauss}},\ }\href {\doibase 10.1017/S0022377896005296} {\bibfield
  {journal} {\bibinfo  {journal} {Journal of Plasma Physics}\ }\textbf
  {\bibinfo {volume} {57}},\ \bibinfo {pages} {83} (\bibinfo {year}
  {1997})}\BibitemShut {NoStop}%
\bibitem [{\citenamefont {Nikulsin}\ \emph {et~al.}(2019)\citenamefont
  {Nikulsin}, \citenamefont {Hoelzl}, \citenamefont {Zocco}, \citenamefont
  {Lackner},\ and\ \citenamefont {G\"unter}}]{nikulsin2019a}%
  \BibitemOpen
  \bibfield  {author} {\bibinfo {author} {\bibfnamefont {N.}~\bibnamefont
  {Nikulsin}}, \bibinfo {author} {\bibfnamefont {M.}~\bibnamefont {Hoelzl}},
  \bibinfo {author} {\bibfnamefont {A.}~\bibnamefont {Zocco}}, \bibinfo
  {author} {\bibfnamefont {K.}~\bibnamefont {Lackner}}, \ and\ \bibinfo
  {author} {\bibfnamefont {S.}~\bibnamefont {G\"unter}},\ }\href {\doibase
  10.1063/1.5122013} {\bibfield  {journal} {\bibinfo  {journal} {Physics of
  Plasmas}\ }\textbf {\bibinfo {volume} {26}},\ \bibinfo {pages} {102109}
  (\bibinfo {year} {2019})}\BibitemShut {NoStop}%
\bibitem [{\citenamefont {Nikulsin}\ \emph {et~al.}(2021)\citenamefont
  {Nikulsin}, \citenamefont {Hoelzl}, \citenamefont {Zocco}, \citenamefont
  {Lackner}, \citenamefont {G\"unter},\ and\ \citenamefont {the
  JOREK~Team}}]{nikulsin2021testing}%
  \BibitemOpen
  \bibfield  {author} {\bibinfo {author} {\bibfnamefont {N.}~\bibnamefont
  {Nikulsin}}, \bibinfo {author} {\bibfnamefont {M.}~\bibnamefont {Hoelzl}},
  \bibinfo {author} {\bibfnamefont {A.}~\bibnamefont {Zocco}}, \bibinfo
  {author} {\bibfnamefont {K.}~\bibnamefont {Lackner}}, \bibinfo {author}
  {\bibfnamefont {S.}~\bibnamefont {G\"unter}}, \ and\ \bibinfo {author}
  {\bibnamefont {the JOREK~Team}},\ }\href {\doibase 10.1017/S0022377821000477}
  {\bibfield  {journal} {\bibinfo  {journal} {Journal of Plasma Physics}\
  }\textbf {\bibinfo {volume} {87}},\ \bibinfo {pages} {855870301} (\bibinfo
  {year} {2021})}\BibitemShut {NoStop}%
\bibitem [{\citenamefont {Dommaschk}(1986)}]{dommaschk1986representations}%
  \BibitemOpen
  \bibfield  {author} {\bibinfo {author} {\bibfnamefont {W.}~\bibnamefont
  {Dommaschk}},\ }\href {\doibase 10.1016/0010-4655(86)90109-8} {\bibfield
  {journal} {\bibinfo  {journal} {Computer Physics Communications}\ }\textbf
  {\bibinfo {volume} {40}},\ \bibinfo {pages} {203} (\bibinfo {year}
  {1986})}\BibitemShut {NoStop}%
\bibitem [{\citenamefont {Drevlak}, \citenamefont {Monticello},\ and\
  \citenamefont {Reiman}(2005)}]{drevlak2005pies}%
  \BibitemOpen
  \bibfield  {author} {\bibinfo {author} {\bibfnamefont {M.}~\bibnamefont
  {Drevlak}}, \bibinfo {author} {\bibfnamefont {D.}~\bibnamefont {Monticello}},
  \ and\ \bibinfo {author} {\bibfnamefont {A.}~\bibnamefont {Reiman}},\ }\href
  {\doibase 10.1088/0029-5515/45/7/022} {\bibfield  {journal} {\bibinfo
  {journal} {Nuclear Fusion}\ }\textbf {\bibinfo {volume} {45}},\ \bibinfo
  {pages} {731} (\bibinfo {year} {2005})}\BibitemShut {NoStop}%
\bibitem [{\citenamefont {Hindenlang}\ \emph {et~al.}(2019)\citenamefont
  {Hindenlang}, \citenamefont {Maj}, \citenamefont {Strumberger}, \citenamefont
  {Rampp},\ and\ \citenamefont {Sonnendr\"ucker}}]{hindenlang2019gvec}%
  \BibitemOpen
  \bibfield  {author} {\bibinfo {author} {\bibfnamefont {F.}~\bibnamefont
  {Hindenlang}}, \bibinfo {author} {\bibfnamefont {O.}~\bibnamefont {Maj}},
  \bibinfo {author} {\bibfnamefont {E.}~\bibnamefont {Strumberger}}, \bibinfo
  {author} {\bibfnamefont {M.}~\bibnamefont {Rampp}}, \ and\ \bibinfo {author}
  {\bibfnamefont {E.}~\bibnamefont {Sonnendr\"ucker}},\ }\href
  {https://hiddensymmetries.princeton.edu/meetings/simons-hour-talks} {\enquote
  {\bibinfo {title} {{GVEC}: A newly developed {3D} ideal {MHD} {G}alerkin
  {V}ariational {E}quilibrium {C}ode},}\ }\bibinfo {howpublished} {Presentation
  given in 'Simons Collaboration on Hidden Symmetries and Fusion Energy'}
  (\bibinfo {year} {2019})\BibitemShut {NoStop}%
\bibitem [{\citenamefont {Hirshman}\ and\ \citenamefont
  {Whitson}(1983)}]{hirshman1983steepest}%
  \BibitemOpen
  \bibfield  {author} {\bibinfo {author} {\bibfnamefont {S.~P.}\ \bibnamefont
  {Hirshman}}\ and\ \bibinfo {author} {\bibfnamefont {J.~C.}\ \bibnamefont
  {Whitson}},\ }\href {\doibase 10.1063/1.864116} {\bibfield  {journal}
  {\bibinfo  {journal} {The Physics of Fluids}\ }\textbf {\bibinfo {volume}
  {26}},\ \bibinfo {pages} {3553} (\bibinfo {year} {1983})}\BibitemShut
  {NoStop}%
\bibitem [{\citenamefont {Hirshman}\ and\ \citenamefont
  {Betancourt}(1991)}]{hirshman1991preconditioned}%
  \BibitemOpen
  \bibfield  {author} {\bibinfo {author} {\bibfnamefont {S.}~\bibnamefont
  {Hirshman}}\ and\ \bibinfo {author} {\bibfnamefont {O.}~\bibnamefont
  {Betancourt}},\ }\href {\doibase
  https://doi.org/10.1016/0021-9991(91)90267-O} {\bibfield  {journal} {\bibinfo
   {journal} {Journal of Computational Physics}\ }\textbf {\bibinfo {volume}
  {96}},\ \bibinfo {pages} {99} (\bibinfo {year} {1991})}\BibitemShut {NoStop}%
\bibitem [{\citenamefont {{W VII-A Team}}(1980)}]{w7a1980stabilization}%
  \BibitemOpen
  \bibfield  {author} {\bibinfo {author} {\bibnamefont {{W VII-A Team}}},\
  }\href {\doibase 10.1088/0029-5515/20/9/008} {\bibfield  {journal} {\bibinfo
  {journal} {Nuclear Fusion}\ }\textbf {\bibinfo {volume} {20}},\ \bibinfo
  {pages} {1093} (\bibinfo {year} {1980})}\BibitemShut {NoStop}%
\bibitem [{\citenamefont {Strumberger}\ and\ \citenamefont
  {Günter}(2016)}]{strumberger2016castor3d}%
  \BibitemOpen
  \bibfield  {author} {\bibinfo {author} {\bibfnamefont {E.}~\bibnamefont
  {Strumberger}}\ and\ \bibinfo {author} {\bibfnamefont {S.}~\bibnamefont
  {Günter}},\ }\href {\doibase 10.1088/0029-5515/57/1/016032} {\bibfield
  {journal} {\bibinfo  {journal} {Nuclear Fusion}\ }\textbf {\bibinfo {volume}
  {57}},\ \bibinfo {pages} {016032} (\bibinfo {year} {2016})}\BibitemShut
  {NoStop}%
\bibitem [{\citenamefont {Strumberger}\ and\ \citenamefont
  {Günter}(2019)}]{strumberger2019linear}%
  \BibitemOpen
  \bibfield  {author} {\bibinfo {author} {\bibfnamefont {E.}~\bibnamefont
  {Strumberger}}\ and\ \bibinfo {author} {\bibfnamefont {S.}~\bibnamefont
  {Günter}},\ }\href {\doibase 10.1088/1741-4326/ab314b} {\bibfield  {journal}
  {\bibinfo  {journal} {Nuclear Fusion}\ }\textbf {\bibinfo {volume} {59}},\
  \bibinfo {pages} {106008} (\bibinfo {year} {2019})}\BibitemShut {NoStop}%
\bibitem [{\citenamefont {Pamela}\ \emph {et~al.}(2020)\citenamefont {Pamela},
  \citenamefont {Bhole}, \citenamefont {Huijsmans}, \citenamefont {Nkonga},
  \citenamefont {Hoelzl}, \citenamefont {Krebs},\ and\ \citenamefont
  {Strumberger}}]{pamela2020extended}%
  \BibitemOpen
  \bibfield  {author} {\bibinfo {author} {\bibfnamefont {S.~J.~P.}\
  \bibnamefont {Pamela}}, \bibinfo {author} {\bibfnamefont {A.}~\bibnamefont
  {Bhole}}, \bibinfo {author} {\bibfnamefont {G.~T.~A.}\ \bibnamefont
  {Huijsmans}}, \bibinfo {author} {\bibfnamefont {B.}~\bibnamefont {Nkonga}},
  \bibinfo {author} {\bibfnamefont {M.}~\bibnamefont {Hoelzl}}, \bibinfo
  {author} {\bibfnamefont {I.}~\bibnamefont {Krebs}}, \ and\ \bibinfo {author}
  {\bibfnamefont {E.}~\bibnamefont {Strumberger}},\ }\href {\doibase
  10.1063/5.0018208} {\bibfield  {journal} {\bibinfo  {journal} {Physics of
  Plasmas}\ }\textbf {\bibinfo {volume} {27}},\ \bibinfo {pages} {102510}
  (\bibinfo {year} {2020})}\BibitemShut {NoStop}%
\bibitem [{\citenamefont {Pamela}, \citenamefont {Huysmans},\ and\
  \citenamefont {Benkadda}(2010)}]{pamela2010influence}%
  \BibitemOpen
  \bibfield  {author} {\bibinfo {author} {\bibfnamefont {S.}~\bibnamefont
  {Pamela}}, \bibinfo {author} {\bibfnamefont {G.}~\bibnamefont {Huysmans}}, \
  and\ \bibinfo {author} {\bibfnamefont {S.}~\bibnamefont {Benkadda}},\ }\href
  {\doibase 10.1088/0741-3335/52/7/075006} {\bibfield  {journal} {\bibinfo
  {journal} {Plasma Physics and Controlled Fusion}\ }\textbf {\bibinfo {volume}
  {52}},\ \bibinfo {pages} {075006} (\bibinfo {year} {2010})}\BibitemShut
  {NoStop}%
\bibitem [{\citenamefont {Finn}, \citenamefont {Cole},\ and\ \citenamefont
  {Brennan}(2019)}]{finn2019real}%
  \BibitemOpen
  \bibfield  {author} {\bibinfo {author} {\bibfnamefont {J.~M.}\ \bibnamefont
  {Finn}}, \bibinfo {author} {\bibfnamefont {A.~J.}\ \bibnamefont {Cole}}, \
  and\ \bibinfo {author} {\bibfnamefont {D.~P.}\ \bibnamefont {Brennan}},\
  }\href {\doibase 10.1063/1.5124490} {\bibfield  {journal} {\bibinfo
  {journal} {Physics of Plasmas}\ }\textbf {\bibinfo {volume} {26}},\ \bibinfo
  {pages} {102505} (\bibinfo {year} {2019})}\BibitemShut {NoStop}%
\bibitem [{\citenamefont {Pamela}, \citenamefont {Huijsmans},\ and\
  \citenamefont {Hoelzl}(2022)}]{pamela2022a}%
  \BibitemOpen
  \bibfield  {author} {\bibinfo {author} {\bibfnamefont {S.}~\bibnamefont
  {Pamela}}, \bibinfo {author} {\bibfnamefont {G.}~\bibnamefont {Huijsmans}}, \
  and\ \bibinfo {author} {\bibfnamefont {M.}~\bibnamefont {Hoelzl}},\ }\href
  {\doibase https://doi.org/10.1016/j.jcp.2022.111101} {\bibfield  {journal}
  {\bibinfo  {journal} {Journal of Computational Physics}\ ,\ \bibinfo {pages}
  {111101}} (\bibinfo {year} {2022})}\BibitemShut {NoStop}%
\bibitem [{\citenamefont {Dudson}\ \emph {et~al.}(2009)\citenamefont {Dudson},
  \citenamefont {Umansky}, \citenamefont {Xu}, \citenamefont {Snyder},\ and\
  \citenamefont {Wilson}}]{dudson2009bout}%
  \BibitemOpen
  \bibfield  {author} {\bibinfo {author} {\bibfnamefont {B.}~\bibnamefont
  {Dudson}}, \bibinfo {author} {\bibfnamefont {M.}~\bibnamefont {Umansky}},
  \bibinfo {author} {\bibfnamefont {X.}~\bibnamefont {Xu}}, \bibinfo {author}
  {\bibfnamefont {P.}~\bibnamefont {Snyder}}, \ and\ \bibinfo {author}
  {\bibfnamefont {H.}~\bibnamefont {Wilson}},\ }\href {\doibase
  https://doi.org/10.1016/j.cpc.2009.03.008} {\bibfield  {journal} {\bibinfo
  {journal} {Computer Physics Communications}\ }\textbf {\bibinfo {volume}
  {180}},\ \bibinfo {pages} {1467} (\bibinfo {year} {2009})}\BibitemShut
  {NoStop}%
\bibitem [{\citenamefont {Dudson}\ \emph {et~al.}(2016)\citenamefont {Dudson},
  \citenamefont {Madsen}, \citenamefont {Omotani}, \citenamefont {Hill},
  \citenamefont {Easy},\ and\ \citenamefont
  {L{\o}iten}}]{dudson2016verification}%
  \BibitemOpen
  \bibfield  {author} {\bibinfo {author} {\bibfnamefont {B.~D.}\ \bibnamefont
  {Dudson}}, \bibinfo {author} {\bibfnamefont {J.}~\bibnamefont {Madsen}},
  \bibinfo {author} {\bibfnamefont {J.}~\bibnamefont {Omotani}}, \bibinfo
  {author} {\bibfnamefont {P.}~\bibnamefont {Hill}}, \bibinfo {author}
  {\bibfnamefont {L.}~\bibnamefont {Easy}}, \ and\ \bibinfo {author}
  {\bibfnamefont {M.}~\bibnamefont {L{\o}iten}},\ }\href {\doibase
  10.1063/1.4953429} {\bibfield  {journal} {\bibinfo  {journal} {Physics of
  Plasmas}\ }\textbf {\bibinfo {volume} {23}},\ \bibinfo {pages} {062303}
  (\bibinfo {year} {2016})}\BibitemShut {NoStop}%
\bibitem [{\citenamefont {Harris}\ \emph {et~al.}(2020)\citenamefont {Harris},
  \citenamefont {Millman}, \citenamefont {van~der Walt}, \citenamefont
  {Gommers}, \citenamefont {Virtanen}, \citenamefont {Cournapeau},
  \citenamefont {Wieser}, \citenamefont {Taylor}, \citenamefont {Berg},
  \citenamefont {Smith}, \citenamefont {Kern}, \citenamefont {Picus},
  \citenamefont {Hoyer}, \citenamefont {van Kerkwijk}, \citenamefont {Brett},
  \citenamefont {Haldane}, \citenamefont {del R{\'{i}}o}, \citenamefont
  {Wiebe}, \citenamefont {Peterson}, \citenamefont {G{\'{e}}rard-Marchant},
  \citenamefont {Sheppard}, \citenamefont {Reddy}, \citenamefont {Weckesser},
  \citenamefont {Abbasi}, \citenamefont {Gohlke},\ and\ \citenamefont
  {Oliphant}}]{harris2020array}%
  \BibitemOpen
  \bibfield  {author} {\bibinfo {author} {\bibfnamefont {C.~R.}\ \bibnamefont
  {Harris}}, \bibinfo {author} {\bibfnamefont {K.~J.}\ \bibnamefont {Millman}},
  \bibinfo {author} {\bibfnamefont {S.~J.}\ \bibnamefont {van~der Walt}},
  \bibinfo {author} {\bibfnamefont {R.}~\bibnamefont {Gommers}}, \bibinfo
  {author} {\bibfnamefont {P.}~\bibnamefont {Virtanen}}, \bibinfo {author}
  {\bibfnamefont {D.}~\bibnamefont {Cournapeau}}, \bibinfo {author}
  {\bibfnamefont {E.}~\bibnamefont {Wieser}}, \bibinfo {author} {\bibfnamefont
  {J.}~\bibnamefont {Taylor}}, \bibinfo {author} {\bibfnamefont
  {S.}~\bibnamefont {Berg}}, \bibinfo {author} {\bibfnamefont {N.~J.}\
  \bibnamefont {Smith}}, \bibinfo {author} {\bibfnamefont {R.}~\bibnamefont
  {Kern}}, \bibinfo {author} {\bibfnamefont {M.}~\bibnamefont {Picus}},
  \bibinfo {author} {\bibfnamefont {S.}~\bibnamefont {Hoyer}}, \bibinfo
  {author} {\bibfnamefont {M.~H.}\ \bibnamefont {van Kerkwijk}}, \bibinfo
  {author} {\bibfnamefont {M.}~\bibnamefont {Brett}}, \bibinfo {author}
  {\bibfnamefont {A.}~\bibnamefont {Haldane}}, \bibinfo {author} {\bibfnamefont
  {J.~F.}\ \bibnamefont {del R{\'{i}}o}}, \bibinfo {author} {\bibfnamefont
  {M.}~\bibnamefont {Wiebe}}, \bibinfo {author} {\bibfnamefont
  {P.}~\bibnamefont {Peterson}}, \bibinfo {author} {\bibfnamefont
  {P.}~\bibnamefont {G{\'{e}}rard-Marchant}}, \bibinfo {author} {\bibfnamefont
  {K.}~\bibnamefont {Sheppard}}, \bibinfo {author} {\bibfnamefont
  {T.}~\bibnamefont {Reddy}}, \bibinfo {author} {\bibfnamefont
  {W.}~\bibnamefont {Weckesser}}, \bibinfo {author} {\bibfnamefont
  {H.}~\bibnamefont {Abbasi}}, \bibinfo {author} {\bibfnamefont
  {C.}~\bibnamefont {Gohlke}}, \ and\ \bibinfo {author} {\bibfnamefont {T.~E.}\
  \bibnamefont {Oliphant}},\ }\href {\doibase 10.1038/s41586-020-2649-2}
  {\bibfield  {journal} {\bibinfo  {journal} {Nature}\ }\textbf {\bibinfo
  {volume} {585}},\ \bibinfo {pages} {357} (\bibinfo {year}
  {2020})}\BibitemShut {NoStop}%
\bibitem [{\citenamefont {Strumberger}\ and\ \citenamefont
  {H{\"o}lzl}(2005)}]{strumberger2005user}%
  \BibitemOpen
  \bibfield  {author} {\bibinfo {author} {\bibfnamefont {E.}~\bibnamefont
  {Strumberger}}\ and\ \bibinfo {author} {\bibfnamefont {M.}~\bibnamefont
  {H{\"o}lzl}},\ }\href {http://hdl.handle.net/11858/00-001M-0000-0027-16AA-6}
  {\enquote {\bibinfo {title} {User manual: Iterative computation of {3D} ideal
  {MHD} equilibria and magnetic fields},}\ }\bibinfo {howpublished} {IPP Report
  5/113} (\bibinfo {year} {2005})\BibitemShut {NoStop}%
\bibitem [{Note1()}]{Note1}%
  \BibitemOpen
  \bibinfo {note} {Time stepping in JOREK involves first linearizing the
  equations in time around the current time step, and then applying applying an
  implicit time advance method, such as the Crank-Nicolson scheme. For more
  details see Ref \protect \rev@citealpnum {nikulsin2021testing}.}\BibitemShut
  {Stop}%
\bibitem [{\citenamefont {Waelbroeck}(1989)}]{waelbroeck1989nonlinear}%
  \BibitemOpen
  \bibfield  {author} {\bibinfo {author} {\bibfnamefont {F.~L.}\ \bibnamefont
  {Waelbroeck}},\ }\href {\doibase 10.1063/1.859165} {\bibfield  {journal}
  {\bibinfo  {journal} {Physics of Fluids B: Plasma Physics}\ }\textbf
  {\bibinfo {volume} {1}},\ \bibinfo {pages} {499} (\bibinfo {year}
  {1989})}\BibitemShut {NoStop}%
\bibitem [{\citenamefont {Huysmans}\ \emph {et~al.}(2001)\citenamefont
  {Huysmans}, \citenamefont {Sharapov}, \citenamefont {Mikhailovskii},\ and\
  \citenamefont {Kerner}}]{huysmans2001modeling}%
  \BibitemOpen
  \bibfield  {author} {\bibinfo {author} {\bibfnamefont {G.~T.~A.}\
  \bibnamefont {Huysmans}}, \bibinfo {author} {\bibfnamefont {S.~E.}\
  \bibnamefont {Sharapov}}, \bibinfo {author} {\bibfnamefont {A.~B.}\
  \bibnamefont {Mikhailovskii}}, \ and\ \bibinfo {author} {\bibfnamefont
  {W.}~\bibnamefont {Kerner}},\ }\href {\doibase 10.1063/1.1398573} {\bibfield
  {journal} {\bibinfo  {journal} {Physics of Plasmas}\ }\textbf {\bibinfo
  {volume} {8}},\ \bibinfo {pages} {4292} (\bibinfo {year} {2001})}\BibitemShut
  {NoStop}%
\bibitem [{\citenamefont {Boyd}(2001)}]{boyd2001chebyshev}%
  \BibitemOpen
  \bibfield  {author} {\bibinfo {author} {\bibfnamefont {J.~P.}\ \bibnamefont
  {Boyd}},\ }\href@noop {} {\emph {\bibinfo {title} {{Chebyshev} and {Fourier}
  Spectral Methods}}},\ \bibinfo {edition} {2nd}\ ed.,\ Dover Books on
  Mathematics\ (\bibinfo  {publisher} {Dover Publications},\ \bibinfo {year}
  {2001})\BibitemShut {NoStop}%
\bibitem [{\citenamefont {Cary}\ and\ \citenamefont
  {Kotschenreuther}(1985)}]{cary1985pressure}%
  \BibitemOpen
  \bibfield  {author} {\bibinfo {author} {\bibfnamefont {J.~R.}\ \bibnamefont
  {Cary}}\ and\ \bibinfo {author} {\bibfnamefont {M.}~\bibnamefont
  {Kotschenreuther}},\ }\href {\doibase 10.1063/1.864973} {\bibfield  {journal}
  {\bibinfo  {journal} {The Physics of Fluids}\ }\textbf {\bibinfo {volume}
  {28}},\ \bibinfo {pages} {1392} (\bibinfo {year} {1985})}\BibitemShut
  {NoStop}%
\bibitem [{\citenamefont {Zanini}\ \emph {et~al.}(2020)\citenamefont {Zanini},
  \citenamefont {Laqua}, \citenamefont {Thomsen}, \citenamefont {Stange},
  \citenamefont {Brandt}, \citenamefont {Braune}, \citenamefont {Brunner},
  \citenamefont {Fuchert}, \citenamefont {Hirsch}, \citenamefont {Knauer},
  \citenamefont {H\"ofel}, \citenamefont {Marsen}, \citenamefont {Pasch},
  \citenamefont {Rahbarnia}, \citenamefont {Schilling}, \citenamefont {Turkin},
  \citenamefont {Wolf},\ and\ \citenamefont {and}}]{zanini2020eccd}%
  \BibitemOpen
  \bibfield  {author} {\bibinfo {author} {\bibfnamefont {M.}~\bibnamefont
  {Zanini}}, \bibinfo {author} {\bibfnamefont {H.}~\bibnamefont {Laqua}},
  \bibinfo {author} {\bibfnamefont {H.}~\bibnamefont {Thomsen}}, \bibinfo
  {author} {\bibfnamefont {T.}~\bibnamefont {Stange}}, \bibinfo {author}
  {\bibfnamefont {C.}~\bibnamefont {Brandt}}, \bibinfo {author} {\bibfnamefont
  {H.}~\bibnamefont {Braune}}, \bibinfo {author} {\bibfnamefont
  {K.}~\bibnamefont {Brunner}}, \bibinfo {author} {\bibfnamefont
  {G.}~\bibnamefont {Fuchert}}, \bibinfo {author} {\bibfnamefont
  {M.}~\bibnamefont {Hirsch}}, \bibinfo {author} {\bibfnamefont
  {J.}~\bibnamefont {Knauer}}, \bibinfo {author} {\bibfnamefont
  {U.}~\bibnamefont {H\"ofel}}, \bibinfo {author} {\bibfnamefont
  {S.}~\bibnamefont {Marsen}}, \bibinfo {author} {\bibfnamefont
  {E.}~\bibnamefont {Pasch}}, \bibinfo {author} {\bibfnamefont
  {K.}~\bibnamefont {Rahbarnia}}, \bibinfo {author} {\bibfnamefont
  {J.}~\bibnamefont {Schilling}}, \bibinfo {author} {\bibfnamefont
  {Y.}~\bibnamefont {Turkin}}, \bibinfo {author} {\bibfnamefont
  {R.}~\bibnamefont {Wolf}}, \ and\ \bibinfo {author} {\bibfnamefont {A.~Z.}\
  \bibnamefont {and}},\ }\href {\doibase 10.1088/1741-4326/aba72b} {\bibfield
  {journal} {\bibinfo  {journal} {Nuclear Fusion}\ }\textbf {\bibinfo {volume}
  {60}},\ \bibinfo {pages} {106021} (\bibinfo {year} {2020})}\BibitemShut
  {NoStop}%
\bibitem [{\citenamefont {Strumberger}, \citenamefont {G\"unter},\ and\
  \citenamefont {the Wendelstein 7-X~Team}(2020)}]{strumberger2020linear}%
  \BibitemOpen
  \bibfield  {author} {\bibinfo {author} {\bibfnamefont {E.}~\bibnamefont
  {Strumberger}}, \bibinfo {author} {\bibfnamefont {S.}~\bibnamefont
  {G\"unter}}, \ and\ \bibinfo {author} {\bibnamefont {the Wendelstein
  7-X~Team}},\ }\href@noop {} {\bibfield  {journal} {\bibinfo  {journal}
  {Nuclear Fusion}\ }\textbf {\bibinfo {volume} {60}},\ \bibinfo {pages}
  {106013} (\bibinfo {year} {2020})}\BibitemShut {NoStop}%
\bibitem [{\citenamefont {Yu}\ \emph {et~al.}(2020)\citenamefont {Yu},
  \citenamefont {Strumberger}, \citenamefont {Igochine}, \citenamefont
  {Lackner}, \citenamefont {Laqua}, \citenamefont {Zanini}, \citenamefont
  {Braune}, \citenamefont {Hirsch}, \citenamefont {Höfel}, \citenamefont
  {Marsen}, \citenamefont {Stange}, \citenamefont {Wolf}, \citenamefont
  {Günter},\ and\ \citenamefont {the Wendelstein 7-X~Team}}]{yu2020numerical}%
  \BibitemOpen
  \bibfield  {author} {\bibinfo {author} {\bibfnamefont {Q.}~\bibnamefont
  {Yu}}, \bibinfo {author} {\bibfnamefont {E.}~\bibnamefont {Strumberger}},
  \bibinfo {author} {\bibfnamefont {V.}~\bibnamefont {Igochine}}, \bibinfo
  {author} {\bibfnamefont {K.}~\bibnamefont {Lackner}}, \bibinfo {author}
  {\bibfnamefont {H.}~\bibnamefont {Laqua}}, \bibinfo {author} {\bibfnamefont
  {M.}~\bibnamefont {Zanini}}, \bibinfo {author} {\bibfnamefont
  {H.}~\bibnamefont {Braune}}, \bibinfo {author} {\bibfnamefont
  {M.}~\bibnamefont {Hirsch}}, \bibinfo {author} {\bibfnamefont
  {U.}~\bibnamefont {Höfel}}, \bibinfo {author} {\bibfnamefont
  {S.}~\bibnamefont {Marsen}}, \bibinfo {author} {\bibfnamefont
  {T.}~\bibnamefont {Stange}}, \bibinfo {author} {\bibfnamefont
  {R.}~\bibnamefont {Wolf}}, \bibinfo {author} {\bibfnamefont {S.}~\bibnamefont
  {Günter}}, \ and\ \bibinfo {author} {\bibnamefont {the Wendelstein
  7-X~Team}},\ }\href {\doibase 10.1088/1741-4326/ab9258} {\bibfield  {journal}
  {\bibinfo  {journal} {Nuclear Fusion}\ }\textbf {\bibinfo {volume} {60}},\
  \bibinfo {pages} {076024} (\bibinfo {year} {2020})}\BibitemShut {NoStop}%
\bibitem [{\citenamefont {Haverkort}\ \emph {et~al.}(2016)\citenamefont
  {Haverkort}, \citenamefont {{de Blank}}, \citenamefont {Huysmans},
  \citenamefont {Pratt},\ and\ \citenamefont
  {Koren}}]{haverkort2016implementation}%
  \BibitemOpen
  \bibfield  {author} {\bibinfo {author} {\bibfnamefont {J.}~\bibnamefont
  {Haverkort}}, \bibinfo {author} {\bibfnamefont {H.}~\bibnamefont {{de
  Blank}}}, \bibinfo {author} {\bibfnamefont {G.}~\bibnamefont {Huysmans}},
  \bibinfo {author} {\bibfnamefont {J.}~\bibnamefont {Pratt}}, \ and\ \bibinfo
  {author} {\bibfnamefont {B.}~\bibnamefont {Koren}},\ }\href {\doibase
  10.1016/j.jcp.2016.04.007} {\bibfield  {journal} {\bibinfo  {journal}
  {Journal of Computational Physics}\ }\textbf {\bibinfo {volume} {316}},\
  \bibinfo {pages} {281} (\bibinfo {year} {2016})}\BibitemShut {NoStop}%
\end{thebibliography}%

\end{document}